\documentclass[12pt]{iopart}

\usepackage{epsfig}
\usepackage{ifpdf}
\usepackage{graphicx}

\usepackage{amssymb,lineno,mathbbol,upgreek}

\usepackage[tight]{subfigure}

\usepackage{bbm}
\usepackage{slashed}
\usepackage{calligra}
\DeclareMathAlphabet{\mathcalligra}{T1}{calligra}{m}{n}
\DeclareFontShape{T1}{calligra}{m}{n}{<->s*[2.2]callig15}{}

\usepackage{calligra}
\DeclareMathAlphabet{\mathcalligra}{T1}{calligra}{m}{n}
\DeclareFontShape{T1}{calligra}{m}{n}{<->s*[2.2]callig15}{}
\usepackage{enumitem}

\begin{document}

\title[Vortex solutions and spectra in a weak harmonic trap]{The nonlinear Dirac equation in Bose-Einstein condensates: Vortex solutions and spectra in a weak harmonic trap}

\author{L H Haddad$^1$ and Lincoln D Carr$^{1,2}$}
\address{$^1$Department of Physics, Colorado School of Mines, Golden, CO 80401,USA    \\ $^2$Physikalisches Institut, Universit\"at Heidelberg, D-69120 Heidelberg, Germany}

\ead{\mailto{laith.haddad@gmail.com}, \mailto{lcarr@mines.edu}}

\begin{abstract}
We analyze the vortex solution space of the $(2 +1)$-dimensional nonlinear Dirac equation for bosons in a honeycomb optical lattice at length scales much larger than the lattice spacing. Dirac point relativistic covariance combined with s-wave scattering for bosons leads to a large number of vortex solutions characterized by different functional forms for the internal spin and overall phase of the order parameter. We present a detailed derivation of these solutions which include skyrmions, half-quantum vortices, Mermin-Ho and Anderson-Toulouse vortices for vortex winding $\ell = 1$. For $\ell \ge 2$ we obtain topological as well as non-topological solutions defined by the asymptotic radial dependence. For arbitrary values of $\ell$ the non-topological solutions include bright ring-vortices which explicitly demonstrate the confining effects of the Dirac operator. We arrive at solutions through an asymptotic Bessel series, algebraic closed-forms, and using standard numerical shooting methods. By including a harmonic potential to simulate a finite trap we compute the discrete spectra associated with radially quantized modes. We demonstrate the continuous spectral mapping between the vortex and free particle limits for all of our solutions. 
\end{abstract}


\submitto{\NJP}

\maketitle

\section{Introduction}

Vortices appear in physical settings which span a wide range of energy scales and disciplines ranging from hurricane phenomena~\cite{Emanuel1988} and turbulent flow in classical fluids~\cite{Kolmogorov1941}, to superfluidity in $^4\textrm{He}$~\cite{Landau1941} and general quantum fluids~\cite{Vinen2006,Gasenzer2011}, down to the smallest length scales of Bogomol'nyi-Prasad-Sommerfield states in supersymmetric field theories~\cite{shinsuke2009}. Vortices are relevant from a technological standpoint in quantum computing for example~\cite{Ohmi2010}, as well as in more theoretical areas of research such as galactic halos in Bose-Einstein condensate (BEC) theories of dark matter~\cite{Kain2009}.~In BECs vortices as stable rotating solutions of the nonlinear Schr\"odinger equation (NLSE) are ubiquitous~\cite{anderson2010,tsubota2012}. Indeed, the presence of a persistent quantized rotation may be considered a defining property of superfluidity.~Although a variety of types of vortices are possible in atomic condenstates~\cite{Kawaguchi2011}, the Laplacian inherent to the Schr\"odinger Hamiltonian constrains the number of observable vortex structures in the BEC. For instance, a BEC made up of spin-$F$ bosons gives rise to a $(2F+1)$-component order parameter which nevertheless solves a multi-component NLSE. An alternative method of constructing a multicomponent BEC is to add the internal degrees of freedom by placing a condensate in a two-dimensional honeycomb optical lattice~\cite{haddad2009,haddad2011}. Atoms are condensed into the lowest energy Bloch state then translated to a corner of the Brillouin zone using laser assisted Bragg scattering~\cite{haddad2012}. At wavelengths large compared to the lattice constant, the microscopic details of the lattice manifest as an additional SL$(2,\textrm{C})$ spin group symmetry, i.e., a Dirac point structure emerges~\cite{haddad2009}.

\begin{figure}[t]
\centering
\subfigure{
\label{fig:ex3-a}
 \includegraphics[width=.65\textwidth]{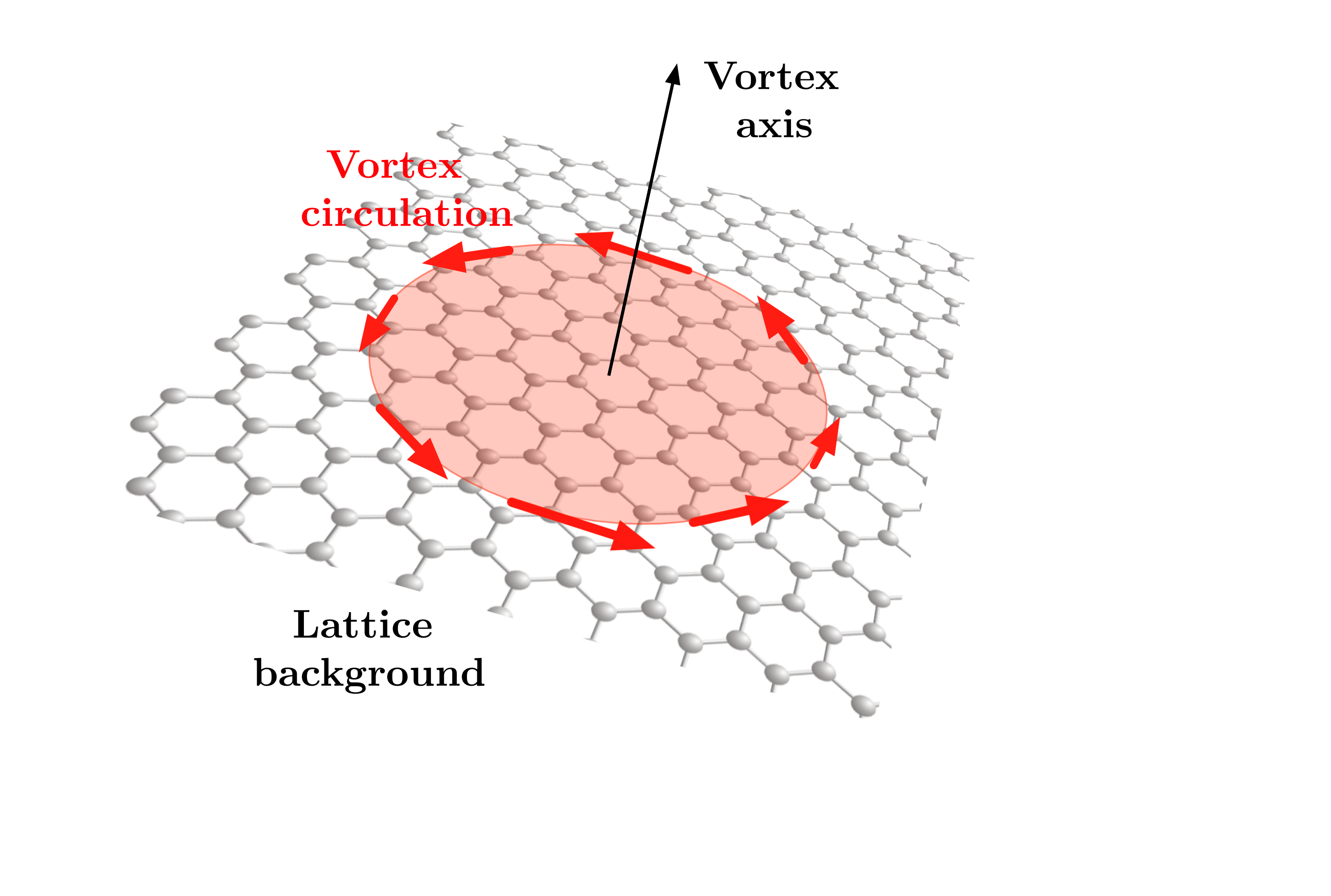}}  \\
\caption[]{(color online) \emph{Relativistic vortex in the honeycomb lattice}. The vortex current is depicted by red arrows with the core towards the interior of the red circle. The relativistic current is quantized around the core and incorporates the overall U(1) phase into the internal SU(2) rotation. }
\label{VortexSchematic}
\end{figure}

 In a previous paper we derived a BEC version of the nonlinear Dirac equation (NLDE) for the case of weak interparticle interactions~\cite{haddad2009}. This has attracted attention from diverse fields of research~\cite{Park2009,Block2010,Szameit2011,Cooper2010,Bahat2010,Dellar2011,Ablowitz2010,Ablowitz2012,Ablowitz2013,Merkl2010,CB2010,Chen2011,Kapit2011,Zhang2009,Parwani2008,Gupta2010,Parwani2011,Kane2007,Kane2010,Buljan2005,Efremidis2003,Peleg2007,Bahat2008,Segev2010,Christodoulides1988,Segev2000,Bittner2010}. Solutions of the NLDE are effectively long wavelength limit lattice envelopes, and so can cover a large number of sites. The healing length in our effectively quasi-2D system is typically about 10 times the lattice constant, and we require the healing length to be small compared to the size of the cylindrical container. In typical experiments, the number of lattice sites is on the order of 100 in a linear direction, thus vortex solutions of the NLDE may be described accurately. In this article we study \emph{relativistic quantum vortices} in the superfluid phase of a Bose gas at the Dirac point of a honeycomb optical lattice focusing on explicit vortex solutions of the NLDE. Our vortices are solutions of equations reminiscent of relativistic systems, but pertain here to ultracold atomic gases, where we work in the mean-field limit throughout. Solution types differ by asymptotic conditions on the amplitude for radius $r$ much greater than the healing length, far from a vortex core, or $r$ much less than the healing length deep inside a vortex core. A general schematic for vortices in our problem is shown in Fig.~\ref{VortexSchematic}. Solutions may also differ by use of the internal degree of freedom in the spinor structure and with respect to quantization of rotation. However, we find that spinor components within a particular solution always differ by one unit of rotation. 
 
 Quantized rotation in a $(2+1)$-dimensional spin-1/2 order parameter is best characterized in terms of the topology of the circulation or phase winding around a singular core. Consider an overall macroscopic phase which wraps around the circular boundary consisting of radial lengths much larger than the healing length. When the internal spin degrees of freedom remain topologically unconstrained we simply referred to this as a \emph{vortex}. In contrast, when the wrapping around the circular boundary includes the internal spin degrees of freedom in a nontrivial way the order parameter is usually called a \emph{texture} or \emph{skyrmion}~\cite{Schroer1996,Kawaguchi2011}. Throughout our work we adhere to the case where the Berry phase is spliced into the overall U(1) symmetry upon condensation of the order parameter. Thus, we do not consider solutions with a macroscopic $\pi$-phase disinclination. We categorized our solutions according to their topological properties based on symmetries of the order parameter manifold. These symmetries must take into account the internal spin degrees of freedom as well as the overall U(1) phase. In particular, localized solutions may asymptotically approach either a local minima, maxima, or saddle point of the effective potential in the mean-field energy functional. The first leads to topological categorization, whereas the latter two cases are not topologically protected configurations.

\begin{figure}[t]
\centering
\subfigure{
\label{fig:ex3-a}
 \includegraphics[width=.55\textwidth]{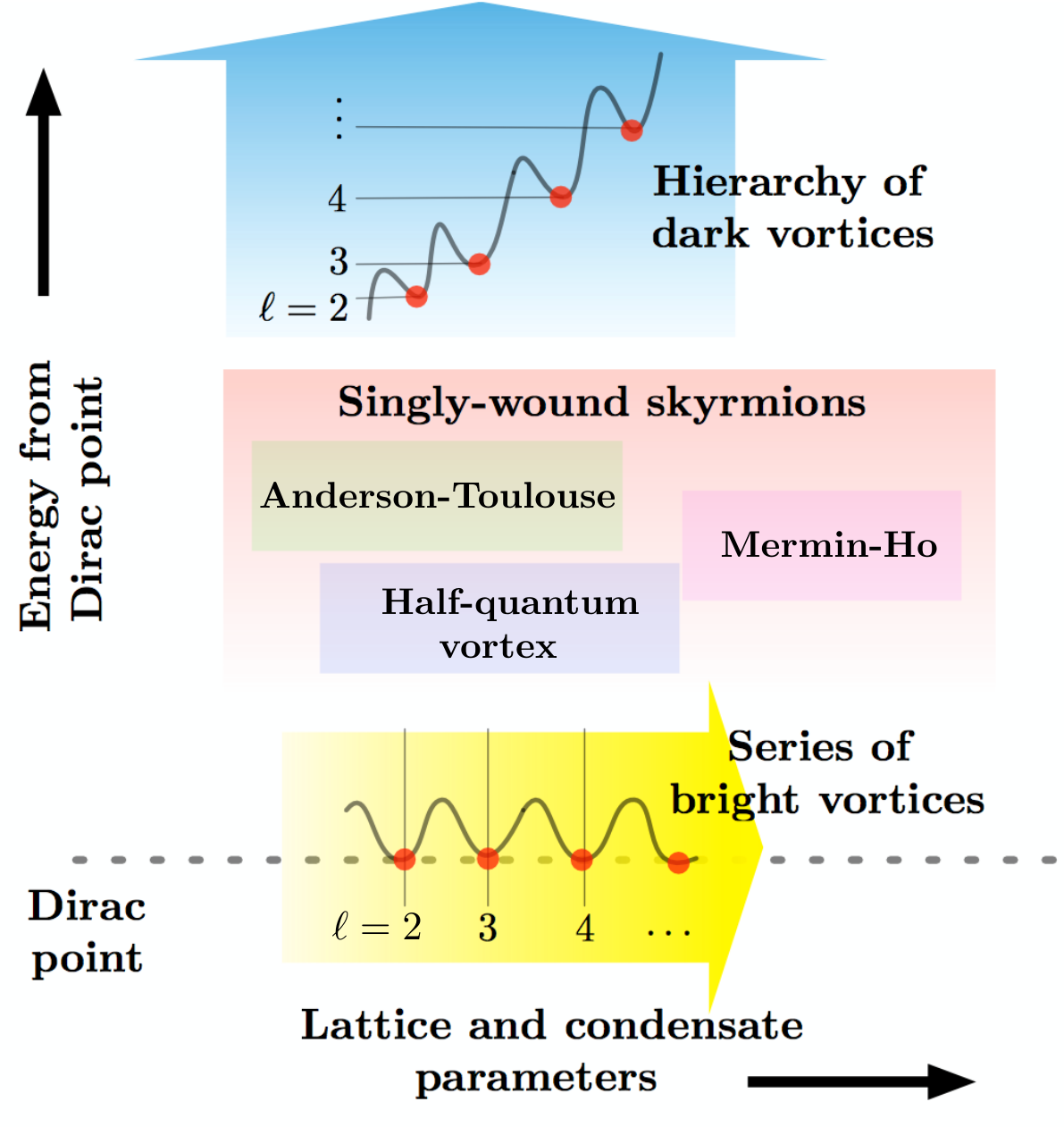}}  \\
\caption[]{(color online) \emph{Schematic of relativistic nonlinear Dirac vortices}. At the Dirac point (horizontal direction) changing the interaction, lattice constant, particle hopping, and density tunes the order parameter between bright vortices (red dots over yellow) with different quantized rotation $\ell \in \mathbb{Z}^+$. The peaks and valleys denote topological transitions between elements of the fundamental group $\pi_1(S^1)$ around the vortex core. Changing the chemical potential away from the Dirac point (vertical direction) transitions the order parameter into skyrmion solutions with a single quantum of rotation $\ell =1$. Higher energy then leads to multiple phase winding states $\ell \ge 2$ (red dots over blue), the energy increasing with winding $\ell$. }
\label{VortexTypes}
\end{figure}

 The presence of the Dirac operator in the NLDE adds new features over conventional spinor BECs. In some cases the derivative terms act to defocus the solution leading to conventional ``dark'' vortices with density minima in an otherwise uniform background. In other cases the cross-mixing of spinor components in the Dirac term has a confining effect resulting in a ``bright'' vortex made of spinning localized density rings over a zero density background. These localized solutions are similar to ones found in ordinary BECs with attractive interactions~\cite{Carr2006,Kevrekidis2012}, although we emphasize that in the present work we consider only repulsive contact interactions. It is significant that our solutions apply to certain spin-orbit coupled BECs~\cite{Stanescu2008}. In the low-energy limit of these theories the linear (Dirac) kinetic term is dominant and one may safely neglect the second order (Schr\"odinger) contribution. Tuning the interactions to eliminate cross terms in the spinor components leads to our NLDE. The various types of NLDE vortices are summarized in Fig~\ref{VortexTypes}.

To put relativistic vortices in a more general context, the large symmetry groups encountered in high energy physics and cosmology admit vortex solutions as fundamental ingredients. Examples include cosmic superstrings~\cite{Polchinski2005,Lee1993,Schaposnik2007}, mirror symmetry of Calabi-Yau manifolds~\cite{Rickles2010,Hori2002}, and brane world scenarios~\cite{Li1998}.~Vortices are generic to gauge theories where spontaneous symmetry breaking occurs. This includes both the Abelian case~\cite{Schaposnik2007,Nielsen1973}, as with Nielsen-Olesen vortices in the Abelian Higgs model, in addition to the non-Abelian case~ \cite{Auzzi2003}, non-Abelian Yang-Mills for example.~Supersymmetric field theories which exhibit weak-strong duality rely fundamentally on the presence of solitons and vortices, as these provide the necessary degrees of freedom to make the dualities possible~\cite{Seiberg1994,Schaposnik2007}.

 Our results are organized as follows. In Sec.~\ref{NLDEVortices}, we express the full time-dependent NLDE in (2+1)D in radial form appropriate for stationary vortices. This reduces the problem to the solution of two coupled first order nonlinear ordinary differential equations. In Sec.~\ref{Symmetries}, symmetries of the order parameter are determined in order to classify the various types of solutions. Section~\ref{Lagrangian} develops the Lagrangian formulation of the nonlinear Dirac equation and connects to the standard theory of spontaneous symmetry breaking. In Sec.~\ref{EnergyFunctional}, we use the mean-field energy functional to examine in detail the interplay between the effective potential and Dirac kinetic terms to gain deeper insight into quasi relativistic vortices. In Sec.~\ref{AnalyticalSolutions}, we solve the NLDE analytically using asymptotic Bessel and algebraic methods. Density and phase plots depicting each solution type are provided. Section~\ref{Reductions} connects our solutions to vortices in the nonlinear Schr\"odinger equation with correction terms, through a semiclassical reduction. In Sec.~\ref{NumericalVortices} we present numerical methods and solutions for vortices. This approach allows us to explore the full vortex landscape by tuning the various physical parameters using the method of numerical shooting. Section~\ref{DiscreteSpectra} presents vortex solutions and spectra in a weak harmonic trapping potential building on results from the previous sections. Finally, in Sec.~\ref{Conclusion} we conclude.

\section{Vortices in the nonlinear Dirac equation}
\label{NLDEVortices}

In this section we express the NLDE in the appropriate coordinates for finding vortex solutions and analyze the solution space at the energy functional level. We show that in different regions of solution space the effective potentials switch between focusing and defocusing forms, providing conditions for either bright (localized ring) or dark (conventional) vortices. The NLDE describes the dynamics of a four-spinor in $(2 +1)$-dimensions, $\Psi \equiv  \left( \Psi_+ , \, \Psi_- \right)^T$, with the upper ($+$) and lower ($-$) two-spinors relating to opposite ${\bf K}$ and ${\bf K}^\prime$ points of the honeycomb lattice and $\Psi : \, \mathbb{R}^3 \to \mathbb{C}^4$. The full NLDE can be expressed as
 \begin{eqnarray}
\left( i \hbar  \gamma^\mu  \partial_\mu   -  U \sum_{i = 0, 1}  \Psi^\dagger \mathrm{M}_i   \Psi   \mathrm{M}_i   \right)    \Psi  = 0   \, ,              \label{NLDE1}
\end{eqnarray} 
where the $4 \times 4$ interaction matrices $\mathrm{M}_i$ are given by 
\begin{eqnarray}
  \mathrm{M}_i \equiv \frac{1}{4} (\gamma^0)^2 + \frac{1}{4} (-1)^{i  + 1/2} \,  \gamma^1 \gamma^2  \, , 
\end{eqnarray}
 constructed to give the correct cubic nonlinearites, local to each spinor component~\cite{haddad2009}. The matrices $\gamma^\mu$ ($\mu = 0, \, 1, \, 2$) are the usual Dirac matrices in the chiral representation~\cite{bjorken64}. Equation~(\ref{NLDE1}) describes a gapless theory which corresponds to massless interacting Dirac spinors. Since the interactions do not couple different spinor components, Eq.~(\ref{NLDE1}) can be split into two sets of equations (one for each of the ${\bf K}$ and ${\bf K}^\prime$ points) of the form
\begin{eqnarray}
i \hbar\,  \partial_t \Psi_+ = \left(   - i \hbar c_l  \, \sigma \cdot \nabla  + U  \sum_{i = 0 , 1}  \Psi_+^\dagger \mathrm{M}^{\mathrm{diag}}_i  \,  \Psi_+  \mathrm{M}^{\mathrm{diag}}_i        \right)  \,  \Psi_+ \, ,  \label{compactNLDE}
\end{eqnarray}
where we use the Pauli matrices $\sigma \equiv ( \sigma_x , \, \sigma_y, \, \sigma_z )$, and $\mathrm{M}^{\mathrm{diag}}_i$ is the nonzero $2 \times 2$ submatrix along the diagonal of $\mathrm{M}_i$ given explicitly by 
\begin{eqnarray}
\mathrm{M}^{\mathrm{diag}}_i \equiv  \frac{1}{4} (\sigma_z)^2 + \frac{1}{4} (-1)^i   \sigma_z  \,. 
\end{eqnarray}
Note the presence of the effective speed of light $c_l$ and interaction strength $U$. These quantities are defined in terms of the more fundamental lattice and atomic parameters as $c_l = t_h a \sqrt{3}/2 \hbar$ and $U = L_z   g \,  \bar{n}^2 \, 3  \sqrt{3} a^2/8$, where $a$ and $t_h$ are the lattice spacing and hopping energy, and $L_z$, $g$, $\bar{n}$ are the vertical oscillator length, two-body interaction, and average three-dimensional atomic density, respectively. The associated long-wavelength healing length associated with the NLDE is then found to be $\xi_\mathrm{Dirac} = t_h a \sqrt{3}/2 U$. For a complete list of the relevant renormalized physical parameters with experimental values see~\cite{haddad2012}.

We seek planar stationary solutions to Eq.~(\ref{compactNLDE}), and thus we require the upper two-spinor to be expressed in factorized form $\Psi_+({\bf r}, t ) =  \mathrm{exp}( - i \mu t/\hbar) \left[  \psi_A({\bf r}), \,      \psi_B({\bf r})     \right]^T$, where $\mu$ is the chemical potential of the system. The spinor components $\psi_{A(B)}({\bf r})$ here are complex functions on the plane, i.e., $\psi_{A(B)}: \, \mathbb{R}^2 \to \mathbb{C}$. For stationary solutions in rectangular coordinates, Eq.~(\ref{compactNLDE}) further reduces to 
 \begin{eqnarray}
      - i \hbar  c_l \left( \partial_x   -  i \partial_y \right) \psi_B(x, y)  +  U \left| \psi_A(x, y) \right|^2 \psi_A(x, y)   =  \mu   \psi_A(x, y)  \label{eqn:CondPsi5} \, , \\
 - i \hbar c_l \left( \partial_x   + i \partial_y \right) \psi_A(x, y)  +  U \left| \psi_B(x, y)  \right|^2 \psi_B(x, y)   =   \mu     \psi_B(x, y)  \label{eqn:CondPsi6}  \, , 
\end{eqnarray}
with the full solution expressed as a linear combination of solutions from Dirac points $\textbf{K}$ and $\textbf{K}^\prime$. To obtain cylindrically symmetric solutions with arbitrary integer phase winding, i.e. vortices, we transform to plane-polar coordinates. The time-independent spinor wavefunctions are then written as 
\begin{eqnarray}
\psi_A(r, \theta) \, = \,  \pm\,  i \, e^{ i (\ell -1) \theta}   f_A(r) \;  , \;\;\;\;\; \psi_B(r, \theta)  =    e^{ i \ell  \theta }  f_B(r) \, , \label{radialansatz}
\end{eqnarray}
where $\ell$ is the angular momentum quantum number (integral phase winding) and the radial functions $f_{A(B)}$ are real functions of $r$. The asymmetric dependence on the winding number $\ell$ between spinor components in Eq.~(\ref{radialansatz}) is required. In polar coordinates, the derivatives in Eq.~(\ref{eqn:CondPsi5})-(\ref{eqn:CondPsi6}) acquire factors of $\exp(-i \theta)$ and $\exp(i \theta)$, respectively. An offset factor of $\exp(-i \theta)$ in $\psi_A$ is then required in order to separate the rotational and radial dependence. Applying the method of separation of variables forces the particular form of angular dependence in Eq.~(\ref{radialansatz}). Equations.~(\ref{eqn:CondPsi5})-(\ref{eqn:CondPsi6}) then reduce to the two coupled radial equations
\begin{eqnarray}
- \hbar c_l \! \left(\! \partial_{r } +  \frac{\ell}{r} \right)\! f_B(r) + U \left|f_A(r)\right|^2\! f_A(r)    = \mu  f_A(r) \label{eqn:CondPsi7} \, ,  \\
  \hbar c_l\! \left( \!\partial_{r }+ \frac{1\! - \! \ell }{r} \right)\! f_A(r) + U  \left|f_B(r)\right|^2 \! f_B(r)  =   \mu f_B(r)  \label{eqn:CondPsi8}\,  . 
\end{eqnarray}
Equations~(\ref{eqn:CondPsi7})-(\ref{eqn:CondPsi8}) comprise the key system of coupled nonlinear equations of this article. In the sections that follow we will solve these using combinations of analytical and numerical techniques for various asymptotic and core boundary conditions. Localized solutions of Eqs.~(\ref{eqn:CondPsi7})-(\ref{eqn:CondPsi8}) may be categorized by their asymptotic forms, i.e., whether the amplitudes $f_{A(B)}$ have zero or nonzero limits far from the core, a relevant distinction in our problem. In particular, we note that in conventional BECs with repulsive interactions the nonlinearity is always defocusing which leads to a vortex with the familiar constant ambient profile that vanishes only at the core. This contrasts radically to the physics in our problem, which we will see allows for (bipartite) vortices with nonzero asymptotic profiles in addition to ones with asymptotically vanishing  profiles. The latter structures are the bright vortices or ring-vortices previously mentioned.

 \section{Symmetries of the order parameter}
 \label{Symmetries}

 To gain a foundational understanding of vortex solutions of the NLDE, we study first the symmetries of the order parameter. Condensed bosons near the Dirac point of a honeycomb lattice are characterized by U(1) symmetry breaking in the full many body theory, and transfer from the ground state to the metastable state at the edge of the Brillouin zone edge~\cite{haddad2012}. The order parameter is comprised of 2 complex amplitudes $\psi = \left( \psi_A , \, \psi_B \right)^T$ associated with the bipartite lattice, forming a representation of the two-dimensional special unitary group $\mathrm{SU}(2) \cong \mathrm{U}(1)_{\phi} \otimes \mathrm{Spin}(2)$. The Spin group in 2D forms a double cover of the rotation group $\mathrm{SO}(2) \cong \mathrm{U}(1)$. Spin components $\psi_A$ and $\psi_B$ are coupled through the kinetic part of the single particle Hamiltonian but are independent with respect to onsite interactions. Thus, the full symmetry group $G$ for a Dirac-point BEC that leaves the mean-field energy functional invariant is 
\begin{eqnarray}
G  \cong \mathrm{U}(1)_\phi \otimes \mathrm{U}(1)_S \otimes  \left( \mathbb{Z}_{2 } \right)_S \, ,  \label{symmetries}
 \end{eqnarray}
 where the subscripts $\phi$ and $S$ denote the gauge and spin degrees of freedom, respectively. The 2D representation of $G$ is generated by the $2\times2$ unit and Pauli matrices $\mathbb{1}, \, \sigma_x, \, \sigma_y$, whose Lie group is given by $\mathrm{exp}\left( - i \phi \right)  \times \mathrm{exp}\left(  -i   \,   \hat{n}  \cdot \vec{\bf \sigma }/2\right)$, where $\vec{\bf \sigma } = ( \sigma_x, \, \sigma_y )$, $\hat{n} = ( \mathrm{cos}\theta  , \, \mathrm{sin} \theta  )$, $\alpha \in \mathbb{R}$, $\theta$ is the polar angle, and $\phi$ is the gauge degree of freedom. In matrix form, $G$ can be parameterized as 
 \begin{eqnarray}
    G  \cong    e^{i \phi }    \left(  \begin{array}{cc} 
                         0  & e^{ - i  \theta/2    }   \\
                           e^{i  \theta /2  }  & 0  \end{array} \right) \, , \label{Gmatrix}
\end{eqnarray}
where again $\phi$ is the gauge angle, $\theta$ is the polar angle, and where the discrete cyclic (off-diagonal matrix) part of $G$ is isolated by setting $\phi = \theta = 0$. One sees that an arbitrary order parameter $\psi$, obtained by applying $G$ to a representative spinor $\psi_0$, has an additional discrete symmetry $H$. This is evident by taking $\phi \to \phi + \pi$ and $\theta \to \theta + 2 \pi$ in Eq.~(\ref{Gmatrix}) described by the discrete symmetry $\left( \mathbb{Z}_2\right)_{\phi, S}$, leading us to deduce the order parameter manifold $M = G/H$
\begin{eqnarray}
M  \cong \frac{ \mathrm{U}(1)_\phi \otimes \mathrm{U}(1)_S \otimes  \left( \mathbb{Z}_{2 } \right)_S}{\left( \mathbb{Z}_2\right)_{\phi, S}}    \, .   \label{OPM}
\end{eqnarray}

 For the gapless system considered here, we generally require both sublattices to be occupied, so that a representative standard spinor is given by $\psi_0 = \left( 1, \, 1 \right)^T$. An arbitrary order parameter $\psi$ is obtained by applying the symmetry transformation $G$ to $\psi_0$, which gives
\begin{eqnarray}
\psi =   e^{i (  \phi  - \theta/2 )}    \left(  \begin{array}{c} 
                              1  \\
                           e^{i  \theta } \end{array}   \right)  \equiv e^{i  \phi'}    \left(  \begin{array}{c} 
                              1  \\
                           e^{i  \theta } \end{array}   \right)  \, . \label{generalOP}                           
\end{eqnarray}
Requiring the order parameter to be single valued under spatial rotations places a constraint on the gauge angle. In particular, we require $\phi$ to contain a part that depends on the polar angle by defining $\phi'  = \phi + \theta/2$ and taking $\phi'$ as the new gauge parameter. Note that we could have factored Eq.~(\ref{generalOP}) another way and obtained 
\begin{eqnarray}
\psi =   e^{i (  \phi  + \theta/2 )}    \left(  \begin{array}{c} 
                              e^{- i  \theta }  \\
                           1  \end{array}   \right)    \equiv e^{i  \phi'}    \left(  \begin{array}{c} 
                                e^{ - i  \theta }   \\
                         1  \end{array}   \right) \, , \label{generalOP2}                        
\end{eqnarray}
in which case we would define $\phi'  = \phi + \theta/2$. The discrete choice of redefinition $\phi'  = \phi \pm  \theta/2$ splits the order parameter space into right and left chiral spinor manifolds displayed in Eq.~(\ref{generalOP}) and Eq.~(\ref{generalOP2}), respectively. Imposing the condition of single valuedeness for the order parameter has effectively eliminated the extra discrete symmetry $\left( \mathbb{Z}_2\right)_{\phi, S}$. Finally, computing the fundamental group of $M$ then yields the direct sum $\pi_1\!  \left(M\right)   \cong  \pi_1\!  \left(G\right)   \cong  \mathbb{Z}_\phi \oplus \mathbb{Z}_{2 , S}$.

 \section{Lagrangian formulation and spontaneous symmetry breaking} 
 \label{Lagrangian}

 So far we have focused primarily on the symmetries of the order parameter associated with the Dirac point structure. We now examine the effect of the nonlinearity associated with contact atomic interactions. Before we obtain exact vortex solutions to the nonlinear Dirac equation it is helpful to study the theory at the Lagrangian level in order to gain insight into the interplay of kinetic and interaction terms and to acquire a qualitative sense of the kinds of vortex solutions one should  expect. From a field theory viewpoint, Eq.~(\ref{NLDE1}) comprises the Euler-Lagrange equations to the Lagrangian density
 \begin{eqnarray}
\fl  \mathcal{L} =  i \hbar \bar{n}_\mathrm{2D}  \left( \psi_A^* \partial_t  \psi_{A}  + \psi_B^*   \partial_t  \psi_{B} \right) -   i \hbar \bar{n}_\mathrm{2D} c_l  \left[  \psi_A^* ( \partial_x  -  i \partial_y) \psi_B  +  \psi_B^* ( \partial_x  + i \partial_y) \psi_A \right] \nonumber  \\
  -   \frac{U}{4} (  |\psi_A|^4  +   |\psi_B|^4 )  \, ,   \label{eqn:Lagrangian} 
 \end{eqnarray}
where Eq.~(\ref{NLDE1}) is obtained through the standard prescription 
 \begin{eqnarray} 
\partial_\mu \! \left( \frac{ \partial  \mathcal{L}}{ \partial  \psi_{A,\mu}^*}   \right)  +  \frac{\partial  \mathcal{L}}{ \partial \psi_A^*} = 0 \, .
\end{eqnarray}
 Note that a factor of the two-dimensional average density $\bar{n}_\mathrm{2D}$ appears in Eq.~(\ref{eqn:Lagrangian}), which gives $\mathcal{L}$ the correct units of energy (see Ref.~\cite{HaddadStability2015}). The positive sign on the second term is correct, since we are taking derivatives with respect to the conjugate of the field and its derivatives. The more common approach is to differentiate the Lagrange density with respect to the field and its derivatives and take the conjugate of the resulting equations afterwards. This leads to the same result. Eq.(\ref{eqn:Lagrangian}) describes the dynamics of two self-interacting, scalar fields coupled through the spatial derivative terms, with interaction strength $U$. The corresponding Hamiltonian is derived using the formula
\begin{eqnarray}
 \mathcal{H} = \pi_A \psi_{A,t} + \pi_B \psi_{B,t} -  \mathcal{L} \,  ,
 \end{eqnarray}
 where $\pi_A$ and $\pi_B$ are the canonical momenta associated with $\psi_A$ and $\psi_B$ respectively, defined in the usual way by 
 \begin{eqnarray}
 \pi_j = \frac{\partial  \mathcal{L} }{ \partial \psi_{j,t}} \,  ,
 \end{eqnarray}

 The Lagrangian density in plane polar coordinates reads
  \begin{eqnarray}
 \fl \mathcal{L} =  i \hbar \bar{n}_\mathrm{2D} \left(  \psi_A^*  \partial_t \psi_{A}  +  \psi_B^* \partial_t  \psi_{B}\right) -  i \hbar \bar{n}_\mathrm{2D}c_l \! \left[  \psi_A^*  e^{ i \theta} ( \partial_r  -  i \frac{1}{r}\;  \partial_{\theta} ) \psi_B  +   \psi_B^* e^{- i \theta} ( \partial_r + i  \frac{1}{r}\,   \partial_{\theta} ) \psi_A \right] \nonumber \\
  - \frac{U}{4} (  |\psi_A|^4  +   |\psi_B|^4 ) \,  \label{eqn:Lag2}.
 \end{eqnarray}
 Since we consider only the case of repulsive interaction where $U > 0$, and given the linear derivatives in the Lagrangian density, the condensate can lower its energy by finding favorable combinations of fields and gradients that render the derivative terms negative. This leads to the formation of a relativistic vortex on an otherwise uniform positive energy background. Spontaneous symmetry breaking is exhibited in Eq.~(\ref{eqn:Lag2}) by expressing the field energy as the sum of a uniform contribution equal to the chemical potential of the system $\mu$, plus an additional piece that accounts for the presence of a vortex. Substituting the ansatz $\psi_A(r, t)  = e^{- i \mu t/2 \hbar } v_A(r,t)$ and $\psi_B(r, t)  = e^{-  i \mu t/2 \hbar } v_B(r,t)$ in Eq.~(\ref{eqn:Lag2}), where the factor of 2 accounts for an even sublattice asymptotic particle distribution, we obtain 
 \begin{eqnarray}
 \fl  \mathcal{L}  =    i \hbar \bar{n}_\mathrm{2D} \! \left( v_A^*  \partial_t v_{A}    +   v_B^* \partial_t v_{B} \right) - i \hbar \bar{n}_\mathrm{2D} c_l  \left[  v_A^*  e^{ i \theta} ( \partial_r  -  i\,  \frac{1}{r}\,   \partial_{\theta} ) v_B  +   v_B^* e^{- i \theta} ( \partial_r + i\; \frac{1}{r}\,   \partial_{\theta} ) v_A  \right]  \nonumber \\
\fl + \frac{\mu}{2}  \left( | v_A |^2 +  | v_B|^2  \right)  - \frac{U}{4} (  |v_A|^4  +   |v_B|^4 )\,  . \label{EffectiveAction}
 \end{eqnarray}
 Reading off the effective potential energy density in Eq.~(\ref{EffectiveAction}), we find
\begin{eqnarray}
\mathcal{PE} =  \frac{\mu}{2}  \left( | v_A |^2 +  | v_B|^2  \right)  - \frac{U}{4} (  |v_A|^4  +   |v_B|^4 ) \, \label{eqn:PE} .
\end{eqnarray}
Decomposing the sublattice order parameters in terms of amplitude and phase $v_{A(B)} = \sqrt{\rho_{A(B)}} \,  e^{i \phi_{A(B)}}$, we see that the minima of the effective potential energy are located at four points in spin space with infinite U(1) degeneracy at each point:
\begin{eqnarray}
v_{A}^\mathrm{min} = \pm  \sqrt{\frac{2 \mu }{U}}  \,  e^{i \phi_A} \,  , \;\;\;\; v_{B}^\mathrm{min} = \pm  \sqrt{\frac{2 \mu }{U}}  \,  e^{i \phi_B} \, . 
\end{eqnarray}
Thus, we see that topological solutions are constrained beyond the symmetry classifications presented so far. At ultra low temperatures the nonlinearity arising from local onsite atomic interactions locks the order parameter into a one of four possible spin configurations characterized by different relative internal phases. However, taking onto account the phases $\phi_{A(B)}$, the symmetry group of the NLDE nonlinearity reduces to $\mathrm{U}(1)_{\phi_A} \otimes \mathrm{U}(1)_{\phi_B} \otimes  \mathbb{Z}_{A,B}$, where the discrete transformation simply exchanges $v_A$ and $v_B$ in Eq.~(\ref{eqn:PE}). So far, our analysis shows that topological solutions of the NLDE are possible and should involve mapping of one or both $\mathrm{U}(1)_{A(B)}$ phases to the circle S$^1$ at spatial infinity. When both phases are involved, the vortex core describes tunneling between potential minima $(v_{A}^\mathrm{min}, \, v_{B}^\mathrm{min}) \to - (v_{A}^\mathrm{min}, \, v_{B}^\mathrm{min})$. In contrast, when only a single U(1) mapping is involved, a vortex core describes a single tunneling $(v_{A}^\mathrm{min}, \, v_{B}^\mathrm{min}) \to ( - v_{A}^\mathrm{min}, \, v_{B}^\mathrm{min})$ or   $(  v_{A}^\mathrm{min}, \, - v_{B}^\mathrm{min})$. We note that non topological solutions are also possible. These are solutions that extremize Eq.~(\ref{eqn:Lagrangian}), but asymptotically approach saddle points or local maxima of Eq.~(\ref{eqn:PE}). Such solutions are possible even in the presence of strictly repulsive interactions. This is because of the rich structure contained in the spin-orbit type (kinetic) coupling in Eq.~(\ref{eqn:Lagrangian}), the effects of which we shall examine in further detail.

\section{Energy functional analysis for vortices}
\label{EnergyFunctional}

 To study the interplay between the effective potential and kinetic terms of the NLDE, we would like to map out the energy landscape for a vortex configuration in more detail. Here, we work with the mean-field energy functional for a generic vortex by eliminating the angular dependence and incorporating centripetal terms into the effective potential energy. Assuming the vortex form in Eq.~(\ref{radialansatz}) and taking $\ell =0$ in our analysis, the total energy is given by
  \begin{eqnarray}
 E   &=&   \int_0^R \! dr \, \left[  \hbar \bar{n}_\mathrm{2D}c_l \left( -   f_B'  f_A  +   f_A'  f_B  \right)  + V_{\mathrm{eff}}  \right]   \, ,\label{energyfunctional}
 \end{eqnarray}
 where, 
 \begin{eqnarray}
\fl  V_{\mathrm{eff}}[ f_A, f_B ] &\equiv& \frac{\mu^2}{2U} \left( \frac{U}{\mu} f_A^2 -1\right)^2 +  \frac{\mu^2}{2U} \left( \frac{U}{\mu} f_B^2 -1\right)^2 + \hbar \bar{n}_\mathrm{2D} c_l \, \frac{f_B f_A}{r} - \frac{\mu^2}{U} + \mu   \, ,   \label{eqn:Veff}
\end{eqnarray}
and we regulate the energy by introducing an upper cutoff $R$. Note that radial derivatives are indicated by prime notation. Evidently the $f_A  f_B'$ and $f_B  f_A'$ terms provide an attractive force for configurations where $\mathrm{sgn}(f_A  f_B') >0$ and $\mathrm{sgn}(f_A' f_B) < 0$, respectively. In the $f_A,f_B$-plane, the centripetal term in $V_\mathrm{eff}$ dominates for small $r$ and forms a saddle point at the origin (of the $f_A,f_B$-plane) becoming singular when $r=0$. Because of the saddle-point, points on the $f_A$ and $f_B$ axes have zero potential energy but are unstable. If we want to study solutions that begin at the saddle-point, i.e., $f_A(0)=f_B(0) =0$, or on the $f_B$-axis, say, near the saddle-point, we must include the full contribution from the kinetic terms. In fact, we see that a path along the $f_B$-axis defined by $p(t) \colon = t \, \hat{e}_B$, $ 0< t< t_f$, will change the total energy by $\Delta E  = c_l  t_f \, f_A'(0)$. This can be made arbitrarily large and negative by adjusting the value of $f_A'$ at $r=0$. The $V_{\mathrm{eff}}$-landscape flattens rapidly as $r$ increases from zero, so that it may be possible to attempt a solution for which $f_A(0) =0$, $f_B(0)>0$, $f_A'(0)<0$. 


If we consider finite energy solutions at large $r$, $f_A'$ and $f_B'$ must go to zero and the centripetal term may be neglected. In this limit, we find that $V_\mathrm{eff}$ has nine local extrema, four of which are minima for the total energy $E$ and thus associated with topological solutions. We obtain the following asymptotic (large $r$) critical points for $V_\mathrm{eff}$ in the $f_A,f_B$-plane: $( \sqrt{2 \mu/U}, \, \sqrt{ 2\mu/ U})$ - local minimum; $( \sqrt{2 \mu/U}, 0)$ - saddle point; $(\sqrt{2 \mu/U}, - \sqrt{2 \mu/ U})$ - local minimum; $(-\sqrt{2 \mu/U},\, \sqrt{2 \mu/ U})$ - local minimum; $(-\sqrt{2 \mu/U}, \, 0)$ - saddle point; $(-\sqrt{2 \mu/U}, \, -\sqrt{2 \mu/ U})$ - local minimum; $(0,\,  \sqrt{2 \mu/ U})$ - saddle point; $(0, \, -\sqrt{2 \mu/ U})$ - saddle point; $(0, \, 0)$ - local maximum. For our chosen case study with $\ell =0$, i.e., with vorticity in only of spinor component, we then expect topological solutions with properties $f_A(0) = 0$, $|f_A'(0)| \ne 0$, $|f_B(0)|>\sqrt{2 \mu/U}$, $f_B'(0) = 0$, and $\lim_{r \to \infty} (|f_A|,  |f_B| ) = ( \sqrt{2 \mu/U},   \sqrt{2 \mu/U})$. This behavior describes a Mermin-Ho vortex. Note that more general topological solutions exist for arbitrary winding with rotation in both spinor components which must both vanish at $r=0$, in order to maintain the finite energy requirement. Moreover, we find that several non-topological localized solutions exist as well, with asymptotic limits coinciding with local extrema that do not minimize the potential energy. Such solutions may satisfy the conditions $f_A(0) = 0$, $|f_A'(0)| \ne 0$, $f_B(0) \ne 0$, $f_B'(0) = 0$, and $\lim_{r \to \infty} (f_A ,  f_B ) = ( 0 ,  0 )$, which describes a ring-vortex/soliton. Another possible non-topological solutions occurs for the same initial conditions but with asymptotic behavior $\lim_{r \to \infty} (|f_A|,  f_B ) = (  \sqrt{2 \mu/U} ,  0 )$, which describes the vortex/soliton solution.

The appearance of non-topological solutions that are nevertheless localized is a consequence of the various possible combinations of configurations for the spinor components at the core of a vortex. Some of these result in attractive contribution to the total energy leading to self-trapping of the order parameter, which is particularly dramatic in the case of ring-vortex solutions. A look at the energy functional for Eqs.~(\ref{eqn:CondPsi7})-(\ref{eqn:CondPsi8}) will help us gain insight into this effect. The NLDE energy functional associated with a vortex with arbitrary circulation $\ell$ is
\begin{eqnarray}
\fl E =  \int \! dr \! \left[  \hbar  \bar{n}_\mathrm{2D}c_l \! \left(  f_A'  f_B - f_B'  f_A     + f_B f_A \frac{2 \ell -1 }{r}   \right) + \frac{U}{4} f_A^4 + \frac{U}{4} f_B^4                \right]  \, . \label{NLDEFunct}
\end{eqnarray}
Taking a different viewpoint, we can define the component-specific effective potentials as 
\begin{eqnarray}
\fl U^A_\mathrm{eff} &\equiv&   f_A'  f_B + f_B f_A \frac{2 \ell -1 }{r}    + \frac{U}{4} \left( f_A^4    +   f_B^4  \right) \,  \equiv  \,  f_A'  f_B  + U_\ell + U_{a-a}  \, , \label{Apot} \\ 
\fl U^B_\mathrm{eff} &\equiv&  - f_B'  f_A + f_B f_A \frac{2 \ell -1 }{r}    + \frac{U}{4} \left( f_B^4   +  f_A^4  \right) \,  \equiv  \,  -  f_B'  f_A  + U_\ell + U_{a-a}\, , \label{Bpot}
\end{eqnarray}
where we have introduced condensed notation for the angular momentum and atom-atom interactions $U_\ell, \, U_{a-a}$, respectively. The energy functional Eq.~(\ref{NLDEFunct}) reduces to the form
\begin{eqnarray}
E =   \int \! dr \left[   \hbar  \bar{n}_\mathrm{2D}c_l \left( f_A'  f_B -  f_B' f_A      \right) + U^A_\mathrm{eff} + U^B_\mathrm{eff}          \right]  \,. 
\end{eqnarray}
The effective potentials $U^{A (B)}_\mathrm{eff}$ encapsulate the angular momentum, binary interactions, and derivative-amplitude bilinear terms. These effective potentials exhibit transitions between focusing and defocusing forms for different regions of spinor solution space. A convenient way to characterize such transitions is through the eigenvalue of the 2D massless radial Dirac operator defined as 
\begin{eqnarray}
   D_\ell \equiv \left( \begin{array}{ c c} 
                                         0 &   \left(\partial_r +  \frac{\ell }{r}   \right) \\
                                           - \left(\partial_r +  \frac{ 1- \ell  }{r}   \right) &  0 \\
                                          \end{array}  \right) \, . 
\end{eqnarray}                                
Denoting the continuous spectrum of $D_\ell$ as $\mathit{D}_\ell \equiv \mathrm{Spec}(D_\ell)$ $\subset$ $\left( - \infty , + \infty \right)$, we can further divide the spectrum into the negative, null, and positive subspaces $\left\{ \lambda_-   \,  | \,   \lambda_- \in  \left( \, - \infty \,  , \, 0 \, \right)  \right\}$, $\left\{ \lambda_0  \,  | \,   \lambda_0  =  0    \right\}$, $\left\{ \lambda_+   \,  | \,   \lambda_+ \in  \left( \, 0 \, , +  \infty \,  \right) \right\}$, respectively. The nature of the effective potentials $U^A_\mathrm{eff}$ and $U^B_\mathrm{eff}$ (and thus vortex solutions) is linked to the character of the subspace of $\mathit{D}_\ell$ that dominates a particular region of the NLDE solution space. Table~\ref{solutionspace} displays the various regions based on analysis of the effective potentials in Eqs.~(\ref{Apot})-(\ref{Bpot}). 
\begin{table}[t]
\begin{tabular}{@{   \hspace{-.25pc}   }l  @{\hspace{2pc}} l  @{\hspace{2pc}} l @{\hspace{2pc}} l}                    
 \hline \hline 
 
 Conditions for $f_A$, $f_B$ and $f_A'$, $f_B'$  &  $U^A_\mathrm{eff}$     &  $U^B_\mathrm{eff}$  &   $\mathit{D}_\ell$  \\   
      
    \hline   \vspace{-.5pc}\\ 
    1.   $f_A , \, f_B ,  f_A' , \, f_B'  >0 $;  \, $f_A f_B' > U_\ell + U_{a-a}$;    &          def.    &   foc.     & null  \\
        \hspace{.6pc}  $f_A \gg  f_A'$;  $f_B \ll  f_B'$  &        &          &       \\
    2.   $f_A , \, f_B, \, f_B' > 0$;  $f_A'  < 0 $; \, $f_B |f_A'|, \, f_A f_B' > U_\ell + U_{a-a}$;     &         foc.    &   foc.  &  neg.      \\
 
   \hspace{.6pc} $f_A \gg  |f_A'|$;  $f_B \gg f_B'$       &                   &      &      \\

   3.   $f_A , \, f_B, \, f_B' > 0$;  $f_A'  < 0 $; \, $f_B |f_A'|, \, f_A f_B' > U_\ell + U_{a-a}$;     &         foc.     &   foc.  &  neg.     \\
 
   \hspace{.6pc} $f_B \approx f_A$;  $|f_A'| \approx f_B'$       &                   &      &     \\

    4.   $f_B , \, f_A, \, f_A' > 0$;  $f_B'  < 0 $    &         def.     &   def.  &  pos.  \\

                \hspace{.6pc}      &                   &      &                                    \\

     5.   $f_A , \, f_A'  > 0$;  $ f_B, \, f_B'  < 0 $; \, $|f_B| f_A' < U_\ell + U_{a-a}$;     &         def.     &   def.  &  pos.    \\
 
  \vspace{-.5pc} \\
 \hline \hline

\end{tabular}   
{\caption{\emph{Character of the effective potentials and the Dirac spectrum for different regions of the NLDE solution space.} Effective potentials $U^{A(B)}_\mathrm{eff}$ exhibit transitions between focusing (foc.) and defocusing (def.) forms depending on the values for spinor components and their derivatives. The character of $\mathit{D}$ denotes the dominant contributing elements of the spectrum for given conditions on $f_{A(B)}$ and $f'_{A(B)}$, i.e., the largest contribution to a vortex solution may come from the positive (pos.) or negative (neg.) part of the spectrum or may be an equal admixture of the two (null). Note that conditions 1-5 do not exhaust all possibilities but are listed to demonstrate the variability needed to observe bright and dark vortices. }  \label{solutionspace}}
\end{table}

The conditions in Table~\ref{solutionspace} are not exhaustive but provide some evidence for the simultaneous presence of dark and bright vortices in the same physical system, particularly in the presence of repulsive atomic interactions. Upon obtaining explicit solutions, we will see that condition 1 and 2 describe the upswing and downswing behavior near the apex of the bright vortex and excitations of the asymptotically flat vortex. This reversal in trend for the radial spinor profiles characterizes the self-trappin effect of focusing potentials. Condition 3 applies to all texture solutions whose spinor degrees of freedom are topologically constrained. Condition 4 gives rise to conventional dark vortices with arbitrary phase windings and non-vanishing tails but lacking the topological constraint of the skyrmion.

\section{Analytical vortex solutions}
\label{AnalyticalSolutions}

 In this section we derive explicit vortex solutions of the NLDE using a combination of analytical and numerical techniques. Some of the solutions have been presented in our previous work~\cite{haddad2012} but without the mathematical detail in the present article. Different methods are applicable depending on whether the chemical potential is zero or finite-valued and if the vortex has a single unit of winding or arbitrarily large winding $\ell$. For arbitrary winding and nonzero chemical potential we use both an approximate asymptotic method and numerical shooting. In the case where $\ell =0$ or $1$ exact analytical solutions are possible, and in particular when $\mu =0$ concise algebraic forms occur. For all of the solutions in this section we consider only the case of a spatially infinite condensate. In experiments, this is a good approximation when the healing length, and thus the core of the vortex, is small compared to the oscillator length associated with a trapping potential which sets the size of the BEC.

\subsection{Asymptotic Bessel solutions for large $\ell$}
\label{Asymptotic}

 Returning to the cylindrically symmetric form of the NLDE, Eqs.~(\ref{eqn:CondPsi7})-(\ref{eqn:CondPsi8}), we can see that for winding values $\ell \ge 2$ both $f_A$ and $f_B$ must vanish at $r=0$ due to the presence of the centrifugal terms. Thus, we will treat the special cases $\ell = 0,1$ in a separate section. To obtain vortex solutions we require spatial derivatives to vanish at infinity; this means that for $r  \to \infty$, Eqs.~(\ref{eqn:CondPsi7})-(\ref{eqn:CondPsi8}) yield the asymptotic behavior
 \begin{eqnarray}
 \lim_{\, r \to \infty} f_{A(B)}(r) \left[ |f_{A(B)}(r)|^2 - \mu/U \right]  = 0 \, , 
 \end{eqnarray}
which implies 
\begin{eqnarray}
\lim_{\, r \to \infty} \left| f_{A(B)}(r)\right| = 0 , \; \pm \sqrt{\mu/U } \, .
\end{eqnarray}
 The important point to note here is whether the $f_A(B)$ tends to zero or a non-zero constant as $r \to \infty$. To determine asymptotic forms for the radial functions we first look for exponential decay to zero, or growth or decay to the constants $\pm \sqrt{\mu/U}$, in which case we find no consistent asymptotic solution to Eqs.~(\ref{eqn:CondPsi7})-(\ref{eqn:CondPsi8}). Algebraic decay to zero may be discerned by substituting $f_A(r) =  r^{- \alpha}$ and $f_B(r) =  r^{- \beta}$ into Eqs.~(\ref{eqn:CondPsi7})-(\ref{eqn:CondPsi8}) for $\alpha , \, \beta > 0$, decay or growth to $\pm \sqrt{\mu /U}$ by $f_A(r) =  r^{- \alpha} \pm \sqrt{\mu /U}$ and $f_B(r) =  r^{- \beta} \pm \sqrt{\mu /U}$. For the case of non-vanishing boundary conditions and $\ell \ge 2$ the asymmetry in the angular momentum terms complicates the task of finding solutions. We recall that the factors $\partial_{r }+ \ell / r$  and  $\partial_{r} + (1-\ell)/r$ act as index raising and lowering operators for the Bessel functions $J_n$. One can verify that Bessel functions are nearly exact solutions in the case of a weak nonlinearity given by $U|\Psi|^2/\hbar \bar{n}_\mathrm{2D} c_l \ll 1$. In this limit solutions of Eqs.~(\ref{eqn:CondPsi7})-(\ref{eqn:CondPsi8}) are approximately Bessel functions of the first kind, $J_l$ and $J_{l-1}$.  A weak nonlinearity only slightly modifies the Bessel form by perturbatively scattering $J_l$ and $J_{l-1}$ into higher $\ell$-valued states.

 However, the case of strong nonlinearity is different. In this case we expect the solution to deviate drastically from the Bessel form, especially for the non-vanishing asymptotic profile of a vortex which interests us here. This heuristic discussion motivates a modified Bessel-Fourier expansion as our ansatz, which must include Bessel functions to all orders
 \begin{eqnarray}
 f_A(r)  =  A \, F(r) \,  e^{i Q(r)} \!  \left[ \,  a_0 +   \sum_{n=1}^\infty  a_n  J_n(r)   \right] \,,    \label{eqn:ansatz1}   \\
   f_B(r)  =  (B/A) \, f_A(r)     \, ,                                         \label{eqn:ansatz2} 
\end{eqnarray} 
where $A$ and $B$ are normalization constants, $J_n(r)$ is the Bessel function of the first kind of order $n$, $a_0$ and $a_n$ are the expansion coefficients. Anticipating an asymptotic expansion, we include the real function $Q(r)$ in the complex prefactor in order to cancel oscillations in the Bessel series at large argument. In addition, we have included the radial function $F(r)$, which we require to cancel the $r^{-1}$ asymptotic decay of the Bessel functions. The series that we have chosen to use runs over the Bessel index rather than the usual form where the summation runs over the zeros of a single Bessel function with a fixed index. This choice of expansion is valid but does not offer the convenience of using the standard orthonormal relations for Bessel functions when computing the coefficients $a_n$.

\begin{figure}[t]
\centering
\subfigure{
\label{fig:ex3-a}
 \includegraphics[width = 1 \textwidth]{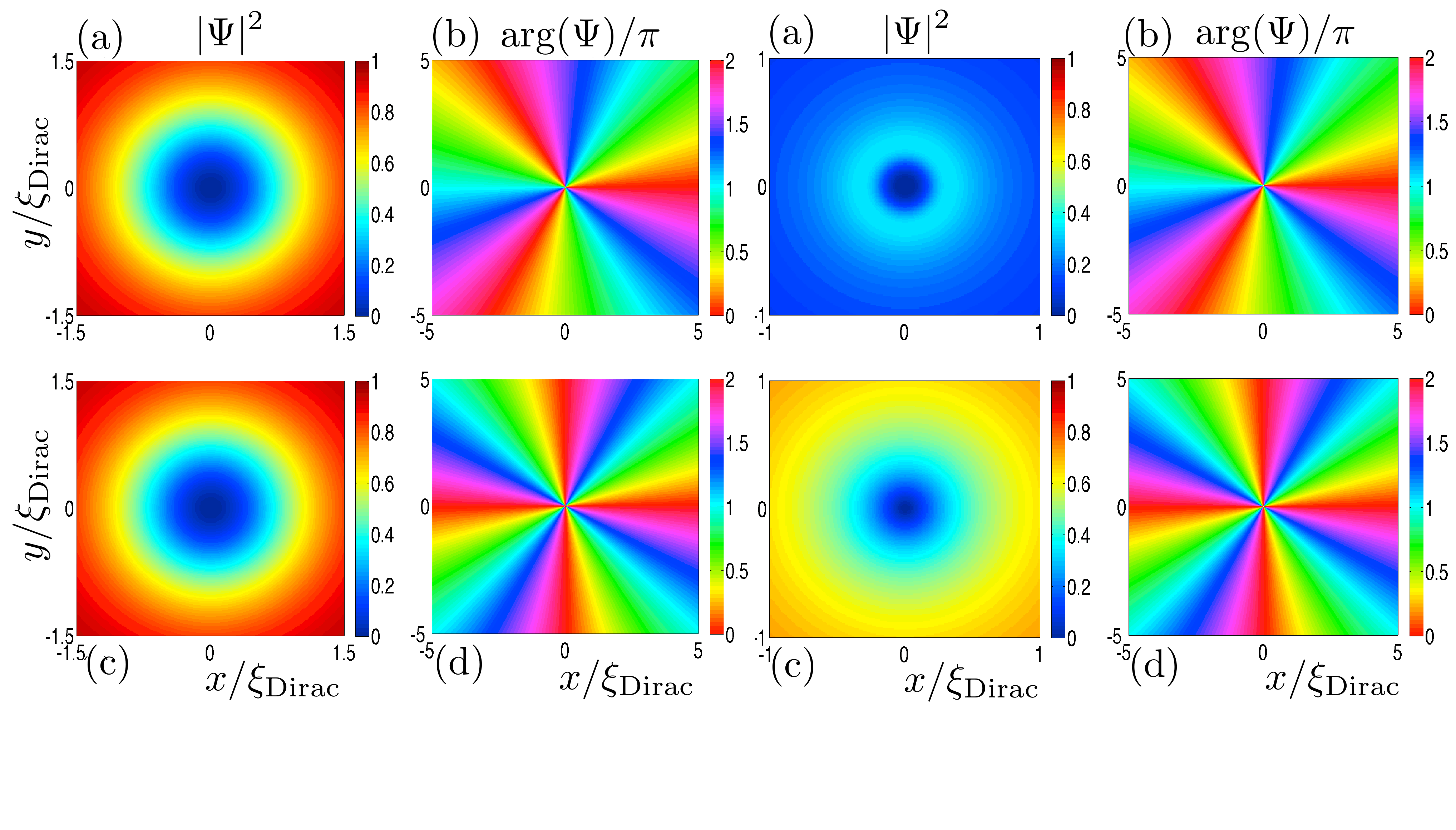}}
\caption[]{(color online) \emph{Phase and density plots for the $\ell =4$ asymptotic Bessel topological vortex (left) and numerical topological vortex (right)}. (a,b) A-sublattice density and phase. (c,d) B-sublattice density and phase. }
\label{DensityPhase1}
\end{figure}

Substituting the ansatz Eqs.~(\ref{eqn:ansatz1})-(\ref{eqn:ansatz2}) into Eq.~(\ref{eqn:CondPsi7})-(\ref{eqn:CondPsi8}), we then consolidate the angular momentum and derivative terms in the resulting series expansion by using the recurrence relations for Bessel functions: $J_n/r\! = \!(J_{n-1} + J_{n+1})/2n$ and $J_n' \!=\! (J_{n-1} - J_{n+1})/2$. We obtain the recursion relations for the coefficients $a_n$ in addition to two first-order differential equation for the functions  $Q$ and $F$
\begin{eqnarray}
   \fl i Q' =    - \frac{F'}{F} -  \frac{ a_0  \ell}{r}  \pm  \frac{ U}{\hbar c_l} i |AF |^2 \left| a_0 + \sum_{n }  a_n  J_n \right|^{2}   \mp i \, \frac{ \mu }{\hbar c_l }\frac{a_0+ \sum_{n }  C a_n  J_n }{a_0+ \sum_{n }  a_n  J_n} \, ,   \label{eqn:reduced1}  \\
  \fl   i Q'  =     - \frac{F'}{F}  +  \frac{ a_0 (1- \ell) }{r}  \pm  \frac{ U}{\hbar c_l} i |AF |^2 \left| a_0 +  \sum_{n }  a_n  J_n \right|^{2}  \mp i \, \frac{ \mu }{\hbar c_l }\frac{a_0+ \sum_{n }  C a_n  J_n }{a_0+ \sum_{n } a_n  J_n} \, ,  \label{eqn:reduced2}  \\
 \fl - \frac{\mu A}{\hbar c_l B} ( 1 - C) a_n  =  \frac{a_{n+1}}{2} \left(\frac{\ell+n+1}{n+1}\right) + \frac{a_{n-1}}{2} \left(\frac{ \ell -n+1}{n-1} \right) \,  ,   \\
\fl \frac{\mu  B}{\hbar c_l  A} ( 1 - C) a_n  =  \frac{a_{n+1}}{2} \left(\frac{2-\ell +n}{n+1}\right) + \frac{a_{n-1}}{2} \left(\frac{ 2-\ell -n}{n-1} \right) \,  .  \label{eqn:recursion} 
\end{eqnarray}
In Eqs.~(\ref{eqn:reduced1})-(\ref{eqn:reduced2}) we have absorbed a fraction $ 1- C < 1 $ from the chemical potential terms into the recursion relations. Solving the recursion relations leads to
\begin{eqnarray}
\hspace{-0pc} A =   \pm\,  i B \;  , \;\;   a_n = \left(\! \pm \, i \frac{ C\,  \mu\,  }{U(2\ell-1)} \,  \!\right)^{n -1} \hspace{-.5pc}  \frac{n \, !}{n^{n-1}}  \, . \label{eqn:expansion}
\end{eqnarray}
Equations~(\ref{eqn:reduced1})-(\ref{eqn:reduced2}) are consistent only in the regime $\ell \gg 1$. Moreover, since $Q$ is a real function we require $F'/F +  a_0  \ell/r = 0$. Choosing $F(r) = r$ leads to $a_0 = -1/\ell$. This choice of $F$  effectively forces the solution to vanish at the origin and also cancels the $r^{-1}$ behavior of the Bessel functions at long distances. For large $\ell$ and $r \gg \xi_{\mathrm{Dirac}}$ in the strong nonlinear regime $\mu/U =1$, Eqs.~(\ref{eqn:reduced1})-(\ref{eqn:reduced2}) read 
\begin{eqnarray}
Q' \simeq   \pm \,  \frac{U}{\hbar c_l}  \left[   |A|^2  r^2  \left|\sum_{n } (-1)^n \left(  a_{2n} \frac{ \mathrm{sin} \, r}{r}  -  a_{2n +1 } \frac{ \mathrm{cos} \, r}{r}         \right)  \right|^{2}            -   1 \right]   \label{Qintegral1} \\
\simeq   \pm \,    \frac{U}{\hbar c_l}  \left[   |A|^2    \left|\sum_{n } (-1)^n \left(  a_{2n} \,  \mathrm{sin} \, r  -  a_{2n +1 } \,  \mathrm{cos} \, r         \right)  \right|^{2}            -   1 \right] \, , \label{Qintegral}
\end{eqnarray}        
which upon integration yields   
\begin{eqnarray}
Q(r) \simeq   \pm \,  \frac{U}{\hbar c_l}  r  \, , 
\end{eqnarray} 
where after integration the oscillating part of Eq.~(\ref{Qintegral}) may be neglected compared to the linear part. Note, that to obtain Eq.~(\ref{Qintegral1}) we used Rayleigh's formula
\begin{eqnarray}
J_n(r) = (-r)^n  \left( \frac{1}{r} \frac{d}{dr} \right)^n \frac{\mathrm{sin}\, r}{r}  \, . 
\end{eqnarray}  
Finally, the parameter $C$ in Eq.~(\ref{eqn:expansion}) must be tuned to a critical value $C_{\textrm{vortex}}$, defined so that the Bessel series converges and is nonzero for all values of $\ell$. We obtain $C_{\textrm{vortex}} =  (5/2) \times (2 \ell -1)$ with closed form solution given by
 \begin{eqnarray}
 f_A(r) =    A  (U r/\hbar c_l) \, e^{ \pm i  ( U r/\hbar c_l)}\!  \sum_{n=1}^\infty   \left(  \frac{5}{2}  \, i   \right)^{n -1} \hspace{-.5pc}  \frac{n \, !}{n^{n-1}}   J_n( \pm U r/\hbar c_l)  \,   .   \label{BesselFinalForm}
 \end{eqnarray}
The overall constant $A$ is determined by normalizing the wavefunction to the particle number. We emphasize here that Eq.~(\ref{BesselFinalForm}) is an approximate solution which applies only very near to or far from the vortex core, and for large winding number. Density and phase plots for the complex Bessel solution are shown in Fig.~\ref{DensityPhase1} on the left with radial profile plots shown in Fig.~\ref{Radial2}(a).

 \subsection{Algebraic solutions}
 
Here we treat solutions for $\ell > 1$ at zero chemical potential $\mu =0$. For zero chemical potential, algebraic closed forms exist for arbitrary values of the winding number. In contrast, for general values of $\mu$ closed forms exist only for $\ell = 0, 1$, which we discuss in the last paragraph of this section. By setting $\mu =0$ in Eqs.~(\ref{eqn:CondPsi7})-(\ref{eqn:CondPsi8}) we obtained exact non-topological vortex solutions. From a technical standpoint eliminating the chemical potential terms in the NLDE simplifies the problem considerably. This is because solutions to the homogeneous non-interacting equations are algebraic forms, not Bessel functions. Thus, vortex solutions do not connect to Bessel functions in the zero-interaction limit. We note here that the precise relationship between the chemical potential $\mu$ and interaction strength $U$ is determined by computing the spectrum of each solution confined within a harmonic trap. This is the subject of Sec.~\ref{DiscreteSpectra} of the present article. To obtain algebraic solutions we start from an ansatz $f_{A} = A r^{\alpha(\beta)}/(1+\gamma \,  r^\delta)^{1/2}$ and likewise for $f_B$ with $A \to B$, where the parameters to be determined are $\alpha, \beta, \gamma,  \delta \in \mathbb{R}$ and $A, B \in \mathbb{C}$ by substitution into the NLDE. Substituting this form for $f_A$ and $f_B$ into Eqs.~(\ref{eqn:CondPsi7})-(\ref{eqn:CondPsi8}) and solving for $A, B, \alpha, \beta, \gamma $ and $\delta$ by matching coefficients of like-power terms, we arrive at
  \begin{eqnarray}
   f_A(r)   =   \frac{A  \;  (U r/\hbar c_l)^{ \ell -1 } }{ \left[ 1 + \frac{|B|^2B}{A(4 \ell -2)}\, ( Ur/\hbar c_l)^{8(\ell-1/2)}   \right]^{ 1/2} }  \, ,  \label{eqn:homogeneous1} \\
    f_B(r)  =  \frac{B  \;  (U r/\hbar c_l)^{3 \ell  - 2} }{ \left[  1 +  \frac{|B|^2B}{A (4\ell -2)}\,  (U r/\hbar c_l)^{8( \ell -1/2)}   \right]^{ 1/2} }    \;  , \label{eqn:homogeneous2} 
 \end{eqnarray}
where it can be proven by construction that no such solution exists when $\mu \ne 0$. Additionally, the constants in Eqs.~(\ref{eqn:homogeneous1})-(\ref{eqn:homogeneous2}) satisfy the constraint and normalization condition 
\begin{eqnarray}
\fl \frac{|A|^2A}{B}  = \frac{|B|^2 B}{C A}  =  4\ell -2  \;  ,  \hspace{2pc}     \int \!  r  dr  \; \frac{ |A|^2 \,   (U r/\hbar c_l)^{2 \ell  -2} + |B|^2 \, (U r/\hbar c_l)^{6 \ell -4}   }{   \left[  1 +  \frac{|B|^2B}{A (4\ell -2)}\,  (U r/\hbar c_l)^{8(\ell -1/2)}   \right]}  = 1 \, .
\end{eqnarray}

\begin{figure}[h]
\centering
\subfigure{
\label{fig:ex3-a}
 \includegraphics[width= 1\textwidth ]{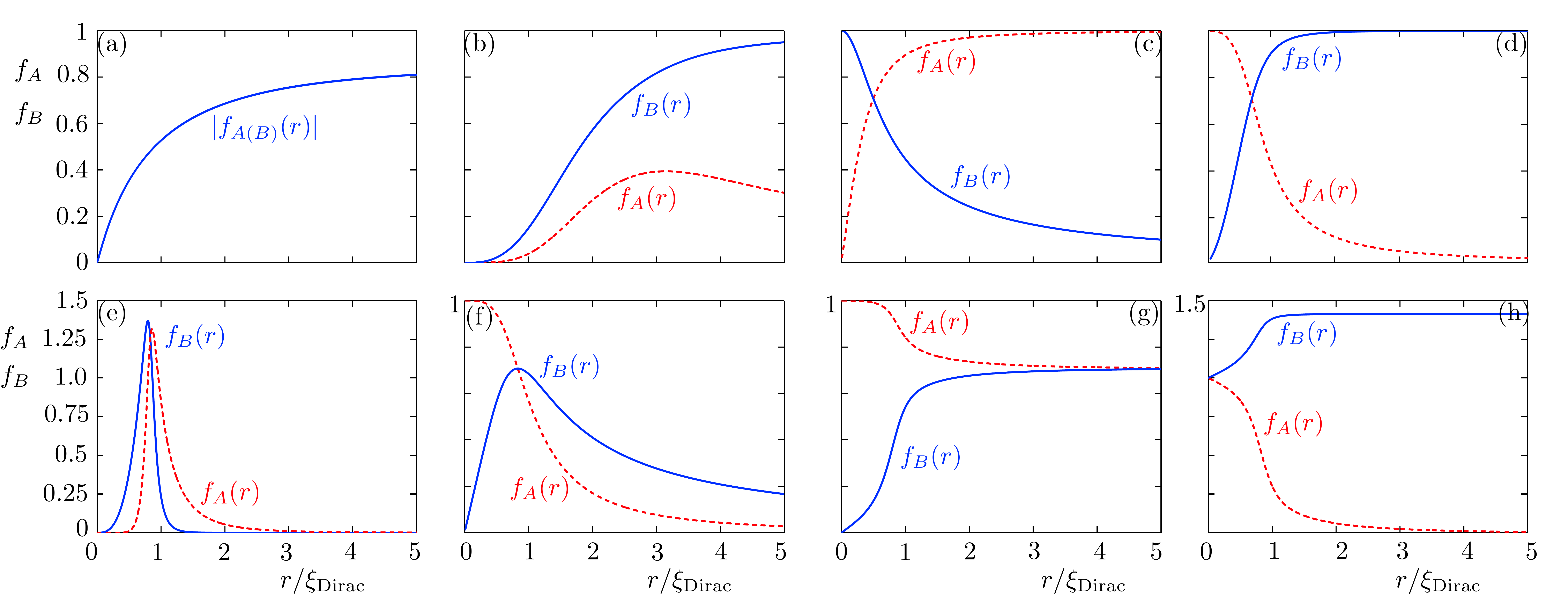}}   \\
\caption[]{(color online) \emph{NLDE vortex radial solutions}. (a) Bessel solution for $\ell=3$; (b) numerical solution for $\ell =4$; (c) vortex/soliton; (d) Anderson-Toulouse vortex; (e) ring-vortex solution for $\ell=4$; (f) ring-vortex/soliton solution; (g) Mermin-Ho vortex; (h) half-quantum vortex. The plot in (a) is the square modulus of the complex asymptotic solution in~Eq.~(\ref{BesselFinalForm}). In each plot, the upper and lower spinor radial solutions are indicated in red and blue, respectively. }
\label{Radial2}
\end{figure}

\begin{figure}[h]
\centering
\subfigure{
\label{fig:ex3-a}
\includegraphics[width = 1 \textwidth]{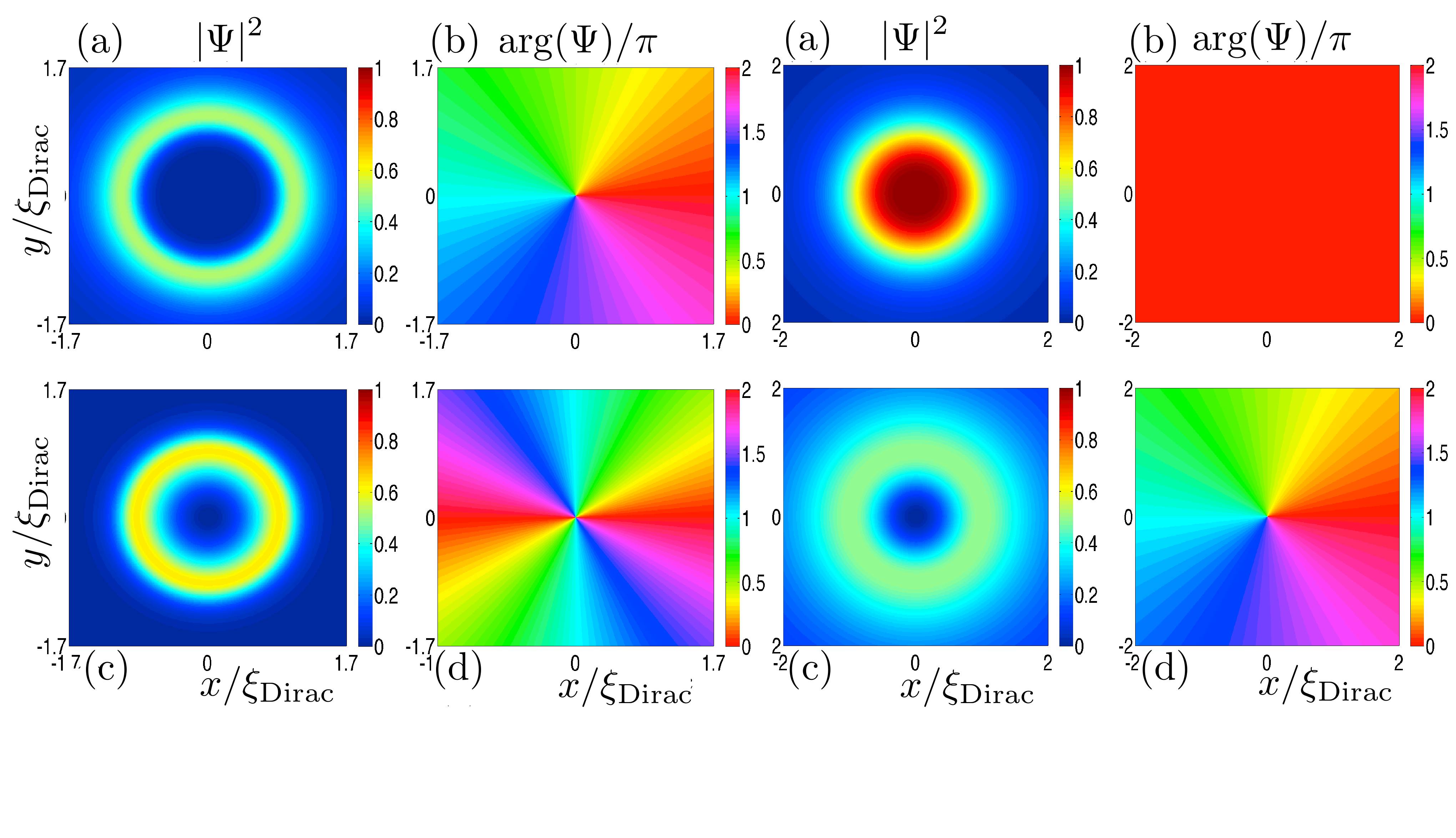}}
\caption[]{(color online) \emph{Phase and density plots for the the $\ell =2$ ring-vortex (left) and the $\ell =1$ ring-vortex/soliton (right)}. (a,b) A-sublattice density and phase. (c,d) B-sublattice density and phase.}
\label{DensityPhase4}
\end{figure}

When the winding number $\ell =0$ or $1$ an angular momentum term appears in only one of Eqs.~(\ref{eqn:CondPsi7})-(\ref{eqn:CondPsi8}). In this case the NLDE is easier to solve and allows for algebraic forms even when $\mu \ne 0$. For the homogeneous case $\mu =0$ though the results in Eqs.~(\ref{eqn:homogeneous1})-(\ref{eqn:homogeneous2}) can be used to obtain the correct solution. Substituting $\ell=1$ into Eqs.~(\ref{eqn:homogeneous1})-(\ref{eqn:homogeneous2}) gives the upper and lower spinor components 
\begin{eqnarray}
 f_A(r)  =  \frac{A  }{\left[ 1 + \frac{|B|^2B}{2A }\,  (U r/\hbar c_l)^{4}   \right]^{ 1/2}  }  \, ,    \label{RVS1}  \\
 f_B(r)  =   \frac{B\, (U r/\hbar c_l)  }{\left[ 1 + \frac{|B|^2B}{2A }\,  (U r/\hbar c_l)^{4}   \right]^{ 1/2} } \, .    \label{RVS2}  
\end{eqnarray}
Equations~(\ref{RVS1})-(\ref{RVS2}) describe a vortex whose density peaks in the shape of a ring with a bright soliton (no rotation) located at its center. This solution corresponds to the ring-vortex/soliton from our previous work~\cite{haddad2011}, but obtained here as a special case of the more general result in Eqs.~(\ref{eqn:homogeneous1})-(\ref{eqn:homogeneous2}). Note that Setting $\ell=0$ in Eqs.~(\ref{eqn:homogeneous1})-(\ref{eqn:homogeneous2}) effectively interchanges the forms for $f_A$ and $f_B$ but leads to the same solution type. The ring-vortex/soliton radial profiles are plotted in Fig.~\ref{Radial2}(d) with the density and phase shown on the right of Fig.~\ref{DensityPhase4}.

Addressing now the case $\mu \ne 0$ and $\ell =1$, an algebraic solution is obtained using an ansatz similar to that used to obtain Eqs.~(\ref{eqn:homogeneous1})-(\ref{eqn:homogeneous2}), which gives
\begin{eqnarray}
 f_A(r) =  \frac{\sqrt{\mu/U} (\mu \, r /\hbar c_l)      }{ \left[ 1 +  (\mu\,  r/\hbar c_l)^2    \right]^{ 1/2} } \, ,        \label{eqn:vortexsoliton1} \\
 f_B(r) = \frac{  \sqrt{\mu /U}  }{ \left[ 1 + (\mu\,  r /\hbar c_l)^2   \right]^{ 1/2} }   \, .   \label{eqn:vortexsoliton2} 
\end{eqnarray}
This solution describes a coreless vortex, a vortex with unit rotation in $f_A$ and a bright soliton in $f_B$ centered at the core of the vortex. This solution can also be obtained by beginning with the ansatz $f_A = \textrm{tanh}[g(r)]$, $f_B= \textrm{sech}[g(r)]$ which, upon substitution into the NLDE, may be directly integrated to give $g(r)= \textrm{arcsinh}( \mu r/\hbar c_l)$; applying a standard identity then yields Eqs.~(\ref{eqn:vortexsoliton1})-(\ref{eqn:vortexsoliton2}). The vortex/soliton radial solution, density and phase are shown in Fig.~\ref{Radial2}(c) and Fig.~\ref{DensityPhase6} (left side), respectively. Radial plots for the $\ell =4$ ring-vortex are shown in Fig.~\ref{Radial2}(c) (left side) with density and phase plots for the case $\ell = 2$ shown in Fig.~\ref{DensityPhase4} (left side).

\subsection{Skyrmion solutions}

\begin{figure}[h]
\centering
\subfigure{
\label{fig:ex3-a}
\includegraphics[width =1 \textwidth]{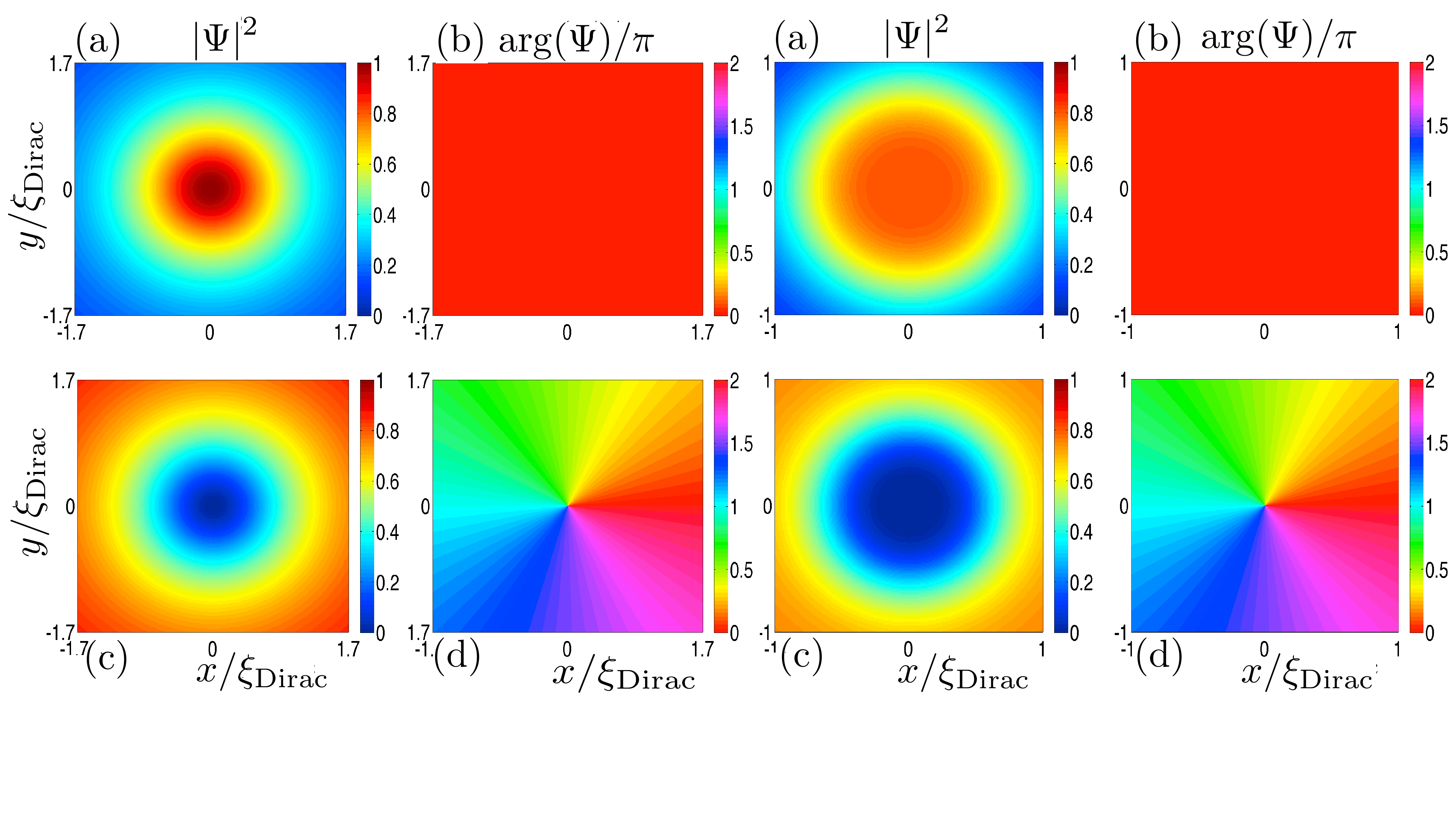}}
\caption[]{(color online) \emph{Phase and density plots for the $\ell =1$ vortex/soliton (left) and $\ell =1$ Anderson-Toulouse skyrmion (right)}. (a,b) A sublattice density and phase. (c,d) B sublattice density and phase.}
\label{DensityPhase6}
\end{figure}

\begin{figure}[h]
\centering
\subfigure{
\label{fig:ex3-a}
 \includegraphics[width = 1\textwidth]{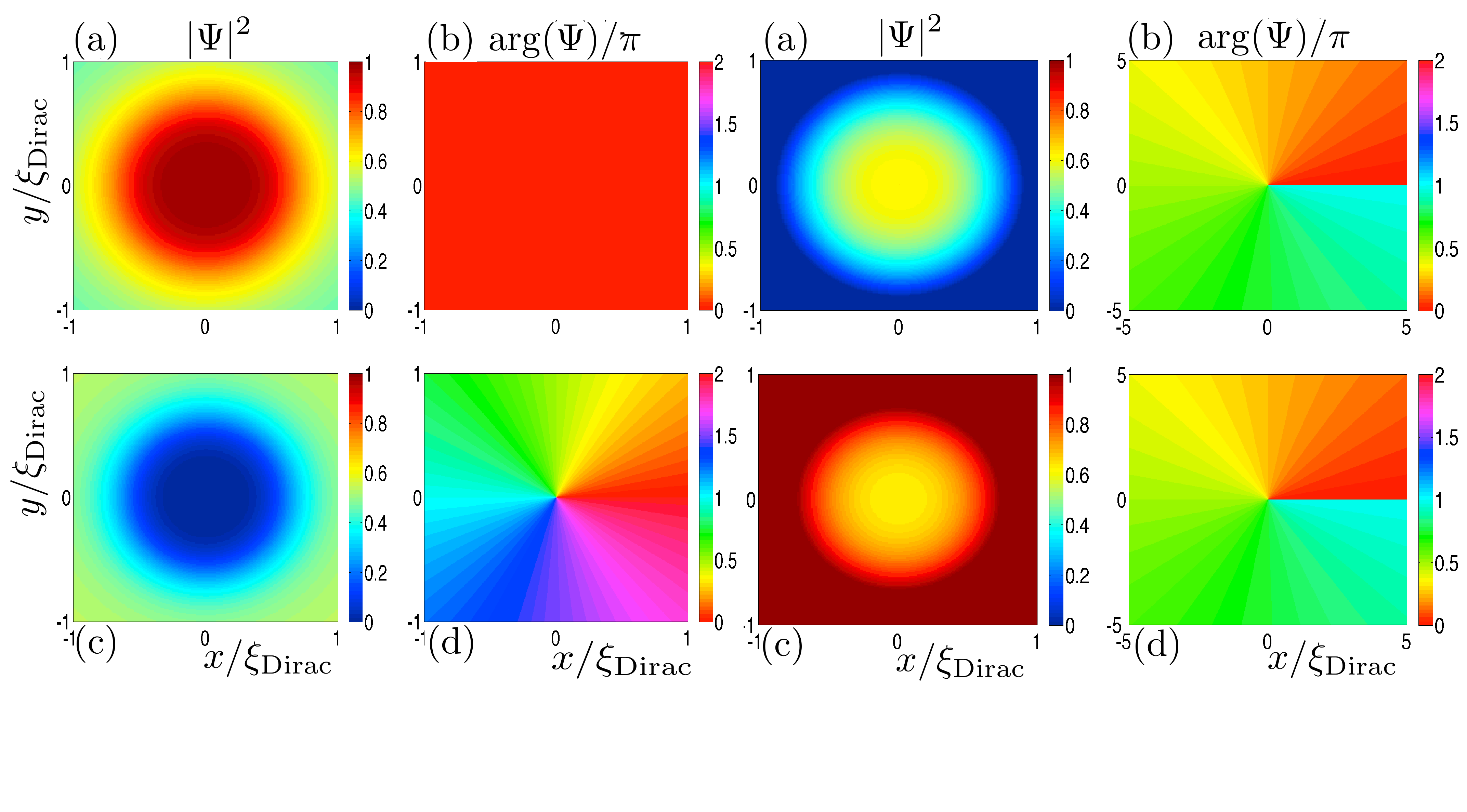}}
\caption[]{(color online) \emph{Phase and density plots for the $\ell =1$ Mermin-Ho skyrmion (left) and half-quantum vortex solutions (right)}. (a,b) A sublattice density and phase. (c,d) B sublattice density and phase.}
\label{DensityPhase8}
\end{figure}

To obtain Skyrmion solutions we choose an ansatz of the form $f_A =   \eta \,  \mathrm{cos}\varphi\,$, $f_B = \eta\, \mathrm{sin} \varphi$, where the parameters $\eta$ and $\varphi$ are functions of the radial coordinate. For background on skyrmions in 2-dimensions see~\cite{Schroer1996,Kasamatsu2005}. Substituting these forms into Eqs.~(\ref{eqn:CondPsi7})-(\ref{eqn:CondPsi8}) reduces the NLDE to two first-order nonlinear ODEs
\begin{eqnarray}
    \frac{d \varphi}{d r}  =  \frac{(1+\ell)}{2r} \mathrm{sin}2 \varphi  + \frac{U}{2 \hbar c_l} \eta^2 (1 + \mathrm{cos}^22 \varphi ) - \frac{\mu}{\hbar c_l}      \, , \label{eqn:ansatzreduced1} \\
    \frac{d \eta}{d r}  =  \eta \frac{\ell}{r} \mathrm{cos}2 \varphi - \eta \frac{1}{r} \mathrm{cos}^2\varphi + \frac{ U}{\hbar c_l} \eta^3\,  \mathrm{sin}4 \varphi  \, . \label{eqn:ansatzreduced2}
 \label{eqn:ansatzreduced2}
\end{eqnarray}
The first point to note is that the centripetal terms place a restriction on the behavior of $\varphi$ for $r \to 0$. Equation~(\ref{eqn:ansatzreduced1}) forces the condition $\varphi \to n \pi /2$ ($n \in \mathbb{Z}$), and the only way to keep Eq.~(\ref{eqn:ansatzreduced2}) finite at $r=0$ is to require $\ell =0, 1$. To satisfy these conditions we find two possible solutions: $\ell =0$ for $\varphi(0)= \pi/2$, and $\ell =1$ for $\varphi(0) = \pi$. Thus, skyrmion solutions exist only for \emph{one unit of angular momentum} in either the upper or the lower two-spinor component. As we noted previously, choosing $\ell =0$ or $1$ simply transfers a unit of rotation from one component to the other. We can reduce the problem of solving Eqs.~(\ref{eqn:ansatzreduced1})-(\ref{eqn:ansatzreduced2}) by further adding the constraint of a constant amplitude $\eta = C$. We then obtain the value for $C$ by examining the asymptotic form of the equations for $r  \rightarrow \infty$. In this limit (assuming finite energy) derivatives vanish and we obtain the asymptotic values
\begin{eqnarray}
\varphi(\infty) = \frac{m \pi}{4} \;  , \;\; \;\;  \eta = \pm \sqrt{\frac{4 \, \mu/U}{ 3 + (-1)^m}}    \; , \label{limitcond}
\end{eqnarray}
where $m \in \mathbb{Z}$. Next, we combine Eqs.~(\ref{eqn:ansatzreduced1})-(\ref{eqn:ansatzreduced2}) into one equation for $\varphi$
\begin{eqnarray}
\hspace{-2pc}  \frac{d \varphi }{d r}  =  \frac{1}{ r} \mathrm{sin}2 \varphi  - \frac{\mu}{\hbar c_l}   + \frac{2 \mu C_m}{\hbar c_l} \left[  1 + \left(     \mathrm{cos}^2 \varphi  -  \frac{4 \mu C_m}{\hbar c_l} \, r\,  \mathrm{sin}4 \varphi \right)^2 \right]  \, , \label{MH}
\end{eqnarray}
for $\ell =1$, and
\begin{eqnarray}
\hspace{-2pc}  \frac{d \varphi }{d r}  =  \frac{1}{ 2 r} \mathrm{sin}2 \varphi  - \frac{\mu}{\hbar c_l}   + \frac{2 \mu C_m}{\hbar c_l} \left[  1 + \left(     \mathrm{sin}^2 \varphi  -  \frac{4 \mu C_m}{\hbar c_l} \, r\,  \mathrm{sin}4 \varphi \right)^2 \right]  \, , \label{AT}
\end{eqnarray}
for $\ell =0$, where $C_m = 1/2$ for odd $m$, and $C_m = 1/4$ for even $m$. Equations.~(\ref{eqn:ansatzreduced1})-(\ref{eqn:ansatzreduced2}) allow for two types of solutions labeled by the subscript $m$; one solution asymptotically approaches $\pi/4$, whereas the other solution approaches $0$. The Anderson-Toulouse solution is obtained for $\varphi(0)=\pi/2$ and $\varphi(\infty)= 0$~\cite{Leonhardt2000,Anderson1977}, whereas the Mermin-Ho solution corresponds to the case $\varphi(\infty)=\pi/4$ in Eq.~(\ref{limitcond})~\cite{Leonhardt2000,Mermin1976}. Note that for the Anderson-Toulouse and Mermin-Ho solutions we find that $\eta=  \sqrt{2\mu/U}$ for the constant envelope factor. The radial profiles for both the Mermin-Ho and Anderson-Toulouse vortices are obtained by solving Eqs.~(\ref{MH})-(\ref{AT}) using a straightforward shooting method~\cite{Carr2006}. We have plotted the radial solutions for both types of skyrmions in Fig.~\ref{Radial2}(d) and (g), with plots of the density and phase shown in Fig.~\ref{DensityPhase6} (right side) and Fig.~\ref{DensityPhase8} (left side).

\subsection{Half-quantum vortices}

The NLDE supports half-quantum vortices, i.e., vortices with half-integer winding $\ell = 1/2$. Such solutions are obtained by forming superpositions of the components $f_A$ and $f_B$ which solve the Mermin-Ho condition~\cite{Salomaa1985,lagoudakis2009}. The necessary requirement is that both spinor components approach the same nonzero value at spatial infinity far from the vortex core (see Fig.~\ref{Radial2}(g)). The appropriate spinor combinations read
\begin{eqnarray}
f_A   =   - i e^{i \theta/2 }   \mathrm{sin} \varphi -   i  e^{-i \theta/2 }  \mathrm{cos} \varphi  \, ,  \label{HQV1} \\
f_B   =   i e^{i \theta/2}  \mathrm{sin} \varphi   -   i e^{-i \theta/2 }  \mathrm{cos} \varphi \, ,  \label{HQV2}
\end{eqnarray}
where $\varphi(r)$ is the Mermin-Ho parameter. Note that Eqs.~(\ref{HQV1})-(\ref{HQV2}) constitute a solution of the full time-dependent NLDE but do not solve the time-independent case. To see that Eqs.~(\ref{HQV1})-(\ref{HQV2}) are associated with a fractional winding number we note that far from the vortex core we have $\varphi \rightarrow \pi/4$, and the wavefunction takes the form 
\begin{eqnarray}
\Psi  =   \, 2 \,\sqrt{\mu/U}\,     e^{-i \theta/2 }    \left[  i\,  \mathrm{cos}(\theta/2) , \;  \mathrm{sin}(\theta/2)   \right]. \label{HQV3}
\end{eqnarray}
From Eq.~(\ref{HQV3}) we compute the geometric phase by circling the vortex core through the Berry phase prescription 
\begin{eqnarray}
 \phi_B \, \equiv \,  \exp{ \left( \oint  \left< \Psi \right| \frac{\partial}{\partial \theta} \left| \Psi \right>  \, d \theta \right) } \, , \label{eqn:berryphase}
\end{eqnarray}
where $\phi_B$ is the Berry phase and $\theta$ is the polar angle. The vortex wavefunction transforms as a Dirac spinor under spatial rotations accumulating a phase factor $\exp{( - i  \sigma_z \theta /2 )}$, where $\sigma_z$ is the third Pauli matrix. Working out the exponential in Eq.~(\ref{eqn:berryphase}) gives
\begin{eqnarray}
  \fl  \int_0^{2\pi} \! d \theta \, e^{ i \theta/2 }    \left[  \begin{array} {c}
           -\, i \,e^{i \theta/2 } \,  \mathrm{cos}( \theta/2)  \; ,\;e^{-i \theta/2 }  \,   \mathrm{sin}(\theta/2)  \,                  \end{array} \right]  \times \frac{\partial}{\partial \theta} \;  e^{ -i \theta/2 }       \left( \begin{array} {c}
             i\, e^{ - i \theta/2 }\, \mathrm{cos}(\theta/2)                    \\
             \, e^{i \theta/2 }\,  \mathrm{sin}(\theta/2)   
\end{array} \right) \nonumber \\
\fl  = \int_0^{2\pi}\! \! \!d \theta  \left( -\,\frac{i}{2}  \right) \nonumber  \\
 \fl  = - i  \, \pi \; . 
\end{eqnarray}
Radial profiles for half-quantum vortex spinor components are plotted in Fig.~\ref{Radial2}(h) where we have taken the polar angle equal to zero. Density and phase plots are shown in Fig.~\ref{DensityPhase8} (right) with general properties of vortices tabulated in Table~\ref{nldevortices}.

 \begin{table*}[phtb]
\begin{center}
\resizebox{16cm}{!}{
\begin{tabular}{  c   c  c   p{2.3cm}  }
\hline   \hline
{\bf Vortex type } & {\bf Winding }  & { \bf Analytic form of} $\Psi({\bf r})$   &  {\bf Asymptotic behavior }    \\
 \hline 
 Vortex/soliton & $\ell =1$  &    $\left[  i \frac{ 1 }{ \sqrt{1 + \, ( r/r_0)^2} } , \, e^{  i \theta}\!  \frac{ (r/ r_0)}{ \sqrt{1 + \, ( r/r_0)^2} } \right]^T$   & \vspace{-1.5pc}$|\psi_A(\infty)|\!=\!1$, $|\psi_B(\infty)|\!=\!0$  \vspace{-.5pc}  \\
Ring-vortex/soliton &$ \ell = 1$  & $\left[  i  \frac{ 1}{ \sqrt{1 + \, ( r/r_0)^4} } , \,  e^{i \theta}\! \frac{ (r/ r_0)}{ \sqrt{1 + \, ( r/r_0)^4} } \right]^T$      &  \vspace{-1.25pc} $|\psi_A(\infty)|\!=\!0$, $| \psi_B(\infty)|\!=\!0$ \\
Anderson-Toulouse skyrmion &$\ell =1$  & $\left[  i \,   \mathrm{cos} \varphi (r/r_0) , \, e^{ i \theta}  \mathrm{sin} \varphi (r/r_0)  \right]^T $ &  $\varphi (\infty)\!=\!0$      \\
 Mermin-Ho skyrmion & $\ell = 1$ & $\left[  i \,  \mathrm{cos} \varphi (r/r_0) , \, e^{ i \theta}  \mathrm{sin} \varphi (r/r_0)  \right]^T$  &  $\varphi (\infty)\!=\!\pi /4$     \\
Half-quantum vortex & $\ell =1$  & $\left[  i  \mathrm{cos}\,  \theta/2 , \;  \mathrm{sin} \, \theta /2  \right]^T $  & $|\Psi (\infty)|\!=\!1$  \\ 
Ring-vortex & $\ell =  2, 3, 4,  ...$ & $\left[  i e^{ i (\ell -1)  \theta}\! \frac{ (r/ r_0)^{ \ell - 1 } }{ \sqrt{1 + \, ( r/r_0)^{8(\ell -1/2)}} } , \,e^{ i  \ell  \theta}\!  \frac{ (r/r_0)^{3 \ell -2 } }{ \sqrt{1 + \, ( r/r_0)^{8(\ell -1/2)}} } \right]^T$  & \vspace{-1.5pc} $| \psi_A(\infty)|\!=\!0$, $| \psi_B(\infty)|\!=\!0$  \\
General topological vortex & $\ell =  2, 3, 4,  ...$&  Numerical  shooting method & $| \psi_A(\infty)|\!=\!1$, $| \psi_B(\infty)|\!=\!1$\\
\hline \hline
\end{tabular} } 
{\caption{\emph{Vortex solutions of the NLDE.} Solutions are described by their phase winding, closed-form expression, and asymptotic properties. The Mermin-Ho, half-quantum vortex, and general topological solutions have conserved topological charge associated with azimuthal phase winding classified according to elements of the fundamental group $\pi_1(S^1) \cong \mathbb{Z}$. In all other cases, the spinor components asymptotically approach either a saddle point or a local maximum of the effective potential. Note that $r_0$ is the length scale associated with the chemical potential or the interaction strength depending on the particular solution. }   \label{nldevortices}}
\end{center}
\end{table*}

 \section{Vortices in the semiclassical reduction to nonlinear Schr\"odinger equation with correction terms}
 \label{Reductions}

We now study the connection of NLDE vortices to their nonlinear Schr\"odinger counterpart. Working again from the ansatz $\psi_{A(B)}({\bf r}, t)  =  e^{-i \mu t/\hbar}  v_{A(B)}({\bf r}, t)$, where we have factored out the dominant energy contribution from the spinor functions, and substituting into the NLDE gives 
\begin{eqnarray}
 \mu   v_{A} +  i  v_{A,t}  =  - i   \mathcal{D}^* v_B + U  |v_A |^2  v_A \label{eqn:2DpsiAw} \, , \\
 \mu  v_{B}  +  i  v_{B,t} =  - i  \mathcal{D}  v_A + U  |v_B |^2  v_B  \label{eqn:2DpsiBw}\,  , 
\end{eqnarray} 
where we have abbreviated the differential operators by using $ \mathcal{D} = \partial_x +i \partial_y $. For clarity of notation, we have set $\hbar = c_l =1$ and used the abbreviated subscript notation to indicate differentiation. We then make the following approximations
 \begin{eqnarray}
 1.\; \;  |v_B |^2\; \approx \;  |v_A |^2\;  \label{eqn:app1}  \\
 2.  \; \;   |v_{B, t} | \; \ll \; |v_{A, x} | \label{eqn:app2} \\
 3.  \; \;  | U | \; \ll \; \mu  \label{eqn:app3}\,  .
 \end{eqnarray} 
With these approximations, Eq.~(\ref{eqn:2DpsiBw}) can be solved for $v_B$ to give  
\begin{eqnarray}
 v_{B}   \approx    - \frac{i}{\mu} (\mathcal{D}  v_A)  \left( 1 +  \frac{U}{\mu} |v_A|^2  \right) \label{eqn:apppsiB} \,  .
\end{eqnarray}
Substituting Eq.~(\ref{eqn:apppsiB}) back into Eq.~(\ref{eqn:2DpsiAw}) 
\begin{eqnarray}
\fl  i  v_{A,t}  =  - i  \mathcal{D}^* \left[ - \frac{i}{\mu} (\mathcal{D} v_A)   \left( 1  + \frac{U}{\mu}  |v_A|^2  \right)    \right]  + U  |v_A |^2  v_A - \mu  v_A \, ,  \\
   \fl  =   - \frac{1}{\mu} (|\mathcal{D}|^2  v_A)  \left( 1  + \frac{U}{\mu}  |v_A|^2   \right)   -  \frac{U}{\mu^2}  (\mathcal{D}v_A ) ( \mathcal{D}^* |v_A|^2)  + U |v_A |^2  v_A  - \mu  v_A \, , \\
   \fl  =    - \frac{1}{\mu} ( {\bf \nabla}^2 v_A)  \left( 1+  \frac{U}{\mu}  |v_A|^2  \right)  -  \frac{U}{\mu^2}  (\mathcal{D}v_A ) ( \mathcal{D}^* |v_A|^2)  + U |v_A |^2 v_A -  \mu v_A\,  . 
 \end{eqnarray}
Note that equality from this point on must be understood within approximations Eqs.~(\ref{eqn:app1})-(\ref{eqn:app3}). Finally we obtain
 \begin{eqnarray}
  \fl    i  v_{A,t} =   - \frac{1}{\mu} {\bf \nabla}^2 v_A  + U |v_A|^2   v_A  -  \mu  v_A - \frac{U}{\mu^2}  \left[ |v_A|^2   ({\bf \nabla}^2 v_A)+ (\mathcal{D}v_A ) ( \mathcal{D}^* |v_A|^2) \right]\,  . \label{eqn:psiAt3}
\end{eqnarray}
Equation~(\ref{eqn:psiAt3}) is the nonlinear Schr\"odinger equation with derivative correction terms and attractive constant background potential $- \mu$. Note that the effective particle mass is $\mu/2$. The correction terms in Eq.~(\ref{eqn:psiAt3}) are smaller than the first three terms on the right hand side due to the extra factor of $U/\mu$ and powers of derivatives of $v_A$. This tells us that localized solutions will differ from standard NLS form only in regions where gradients are steep, i.e., near the core of a vortex.

Next, we look for vortex solutions of Eq.~(\ref{eqn:psiAt3}) by choosing the stationary, axisymmetric factorized form 
\begin{eqnarray}
v_A(r, \theta,t) =  C  e^{- i \omega t}  e^{i n \theta}  v(r) \,  , \label{eqn:psivortex}
\end{eqnarray}
where $C \in \mathbb{R}$ and $n \in \mathbb{Z}$ are constants, and such that 
\begin{eqnarray}
\mathrm{lim}_{ \;r \rightarrow \infty} |v_A| =  C \; \; \Rightarrow \; \;  \mathrm{lim}_{ \;r \rightarrow \infty} |v(r)|  =  1\; \; , \; \; \;  \mathrm{lim}_{ r \rightarrow 0} |v_A| =  0 \,  .
\end{eqnarray}
In plane-polar coordinates, Eq.~(\ref{eqn:psiAt3}) becomes
\begin{eqnarray}
   i  v_{A,t}  =   - \frac{1}{\mu} \left( \frac{\partial^2}{\partial r^2}  +  \frac{1}{r^2}\frac{\partial^2}{\partial \theta^2} +  \frac{1}{r} \frac{\partial }{\partial r}                             \right) v_A + ( U |v_A|^2 - \mu ) v_A  \nonumber   \\
  -  \frac{U}{\mu^2}  |v_A|^2 \left( \frac{\partial^2}{\partial r^2} +  \frac{1}{r^2}\frac{\partial^2}{\partial \theta^2} +  \frac{1}{r} \frac{\partial }{\partial r}   \right)   v_A   \nonumber \\
  -   \frac{U}{\mu^2}  \left[e^{i\theta} \left( \frac{\partial}{\partial r} + i \frac{1}{r} \frac{\partial}{\partial \theta}                             \right)       v_A                                         \right] \left[  e^{-i\theta} \left( \frac{\partial}{\partial r} - i \frac{1}{r} \frac{\partial}{\partial \theta}                             \right)      | v_A|^2                   \right]
 \label{eqn:psiAt6} .
\end{eqnarray}
Inserting Eq.~(\ref{eqn:psivortex}) gives
\begin{eqnarray}
\omega  v  =  - \frac{1}{\mu} \left( \frac{\partial^2}{\partial r^2} -  \frac{n^2}{r^2} + \frac{1}{r} \frac{\partial }{\partial r}                             \right) v + ( U C^2 v^2 - \mu ) v  \nonumber   \\
 -  \frac{U}{\mu^2}  C^2 v^2 \left(  \frac{\partial^2}{\partial r^2} - \frac{n^2}{r^2} + \frac{1}{r} \frac{\partial }{\partial r}   \right)   v    \nonumber \\
  -   \frac{U}{\mu^2} C^2 v  \left[ \left( \frac{\partial}{\partial r} -  \frac{n}{r}   \right)   v  \right] \left[    \left( \frac{\partial}{\partial r}  - \frac{n}{r}   \right) v + \left( \frac{\partial}{\partial r}  +  \frac{n}{r}   \right)       v                                         \right]  .
\end{eqnarray}
Canceling some terms and using condensed notation, we have
\begin{eqnarray}
\omega  v  =  - \frac{1}{\mu} \left( v_{rr} -  \frac{n^2}{r^2} v + \frac{1}{r} v_r    \right)  + ( U C^2 v^2 - \mu )  v  \nonumber   \\
 -  \frac{U}{\mu^2}   C^2 v^2    \left( v_{rr} -  \frac{n^2}{r^2} v+ \frac{1}{r} v_r    \right) -  \frac{U}{\mu^2} C^2 v   \left( v_r - \frac{n}{r} v  \right)     2 v_r \,  .
\end{eqnarray}
More simplifying leads to
\begin{eqnarray}
    \frac{1}{\mu} \left\{ v_{rr} + \frac{1}{r} v_r  + \left[\mu  \left( \mu +  \omega \right) - \frac{n^2}{r^2}\right] v     \right\}   \nonumber   \\
    =   U C^2 v^3 -  \frac{U}{\mu^2}   C^2 v^2    \left( v_{rr}- \frac{n^2}{r^2} v + \frac{1}{r} v_r                       \right) - 2  \frac{U}{\mu^2}   C^2 v  \left( v_r^2 - \frac{n}{r} v v_r \right)      .
\end{eqnarray}
This can be further simplified by making the coordinate change $\xi= \sqrt{\mu \left( \mu + \omega \right)} \,  r $
\begin{eqnarray}
\fl  \left( \mu +  \omega \right) \left[ v_{\xi \xi} +  \frac{1}{\xi} v_\xi  + \left( 1 -  \frac{n^2}{\xi^2}\right) v      \right]   = U C^2 v^3 -  \frac{U}{\mu} \left( \mu  + \omega \right)  C^2 v^2    \left( v_{\xi \xi} - \frac{n^2}{\xi^2} v + \frac{1}{\xi} v_\xi                       \right) \;\nonumber \\
 \fl - 2  \frac{U}{\mu}\left( \mu + \omega \right)  C^2 v   \left( v_\xi^2  - \frac{n}{\xi} v v_\xi \right)  .
\end{eqnarray}
Finally, dividing through by $\left(\mu +  \omega \right)$ gives
\begin{eqnarray}
  \left[ v_{\xi \xi}+  \frac{1}{\xi} v_\xi  + \left( 1 -\frac{n^2}{\xi^2}\right) v      \right]  =   \frac{ U C^2}{\left( \mu +  \omega \right) }\; v^3 - \frac{U}{\mu}   C^2 v^2    \left( v_{\xi \xi}- \frac{n^2}{\xi^2} v +  \frac{1}{\xi} v_\xi                       \right) \nonumber \\
  - 2  \frac{U}{\mu}   C^2 v \left( v_\xi^2 - \frac{n}{\xi} v v_\xi \right)   \label{eqn:bessel}   .
\end{eqnarray}
There are several limits of this equation that are interesting:
\begin{enumerate}

\item  For small $r$ we expect that $v \rightarrow 0$. In this limit the nonlinear terms on the right hand side become negligible so that
\begin{eqnarray}
 \left[ v_{\xi \xi} +   \frac{1}{\xi} v_\xi  +  \left( 1 \; -\; \frac{n^2}{\xi^2}\right) v \   \right]  =  0  \, .
\end{eqnarray}
This is Bessel's equation and the solutions are the well known Bessel functions $J_n(x)$ and $Y_n(x)$. The ones we are interested in are the Bessel functions of the first kind, $J_n(x)$, since these are regular at the origin. Of these, we further restrict solutions to those for which $n >  0$ since these are zero at the origin.

\item  We seek solutions with constant square modulus for large $r$. In this limit, all derivatives may be set to zero in Eq.~(\ref{eqn:bessel}). This allows us to solve for $C$ in terms of the constants $U$, $\mu$, and $\omega$. For large $\xi$ (large $r$),  $v \rightarrow 1$, and Eq.~(\ref{eqn:bessel}) reduces to $1=  U C^2/( \mu + \omega )$, or simply $C =\sqrt{\left( \mu +  \omega \right)/U}$.

\item  For $U\rightarrow 0$, we should retrieve the standard Schr\"odinger result for the radial part of the wavefunction. This is indeed the case since, in this limit, all terms on the right hand side of Eq.~(\ref{eqn:bessel}) vanish and we are left with Bessel's equation as expected. 

\end{enumerate}

 \section{Vortices by the method of numerical shooting}
 \label{NumericalVortices}

\begin{figure}[t]
\centering
\subfigure{
\label{fig:ex3-a}
  \includegraphics[width=.7\textwidth]{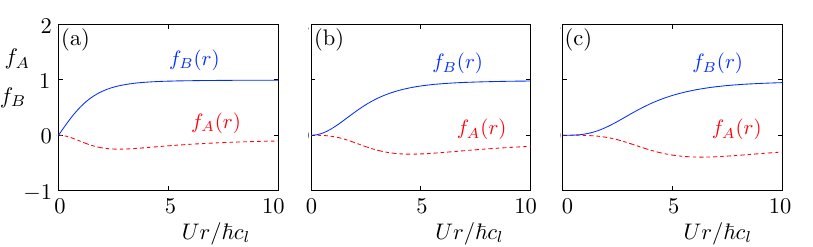}}  \\
\caption[]{(color online) \emph{Radial profiles for $\ell > 1$ vortices by numerical shooting}. (a) Profile for $\ell =2$. (b) Profile for $\ell =3$. (c) Profile for $\ell =4$.}
\label{highervortices}
\end{figure}

\begin{figure}[h]
\centering
\subfigure{
\label{fig:ex3-a}
  \includegraphics[width=.6\textwidth]{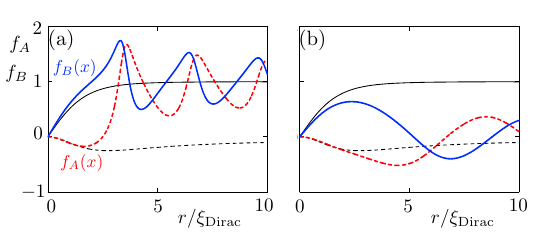}}  \\
\caption[]{(color online) \emph{Convergence of numerical vortex for $\ell=2$}. (a) For $a_1 > a_1^{\textrm{vortex}}$, the solution overshoots to a radial excited state of the vortex. Note that such oscillating solutions converge to unity far from the core. (b) For $a_1 <  a_1^{\textrm{vortex}}$, the solution undershoots and converges to the linear solution Bessel functions. The solid blue and dashed red plots are the A and B sublattice radial wavefunctions, respectively. The solid black and dashed black plots are the exact solutions for the A and B sublattice radial wavefunctions, respectively.}
\label{NumVortexConv}
\end{figure}

In general, vortex profiles for arbitrary phase winding $\ell$ and chemical potential $\mu$ may be obtained using a numerical shooting method as described in Refs.~\cite{Carr2006} and~\cite{haddadcarrsoliton1}. These include topological solutions whose tails do not vanish asymptotically, as well as non-topological vortices where one or both spinor functions have asymptotically vanishing profiles. To proceed we first convert the NLDE to dimensionless form by introducing the rescaled radial coordinate and spinor components
\begin{eqnarray}
\chi \equiv \mu r/(\hbar c_l) \, ,  \;\; \eta_A \equiv \sqrt{U/\mu} f_A \,  , \; \;  \eta_B \equiv \sqrt{U/\mu} f_B  \, ,  \label{rescaling}
\end{eqnarray}
so that Eqs.~(\ref{eqn:CondPsi7})-(\ref{eqn:CondPsi8}) become
\begin{eqnarray}
 \left(\! \partial_{\chi}+\frac{\ell}{\chi} \right)\! \eta_B(\chi) -  \left|\eta_A(\chi)\right|^2\! \eta_A(\chi)     = -  \eta_A(\chi) \, ,  \label{dimensionless1}                       \\
 -   \left( \!\partial_{\chi}+ \frac{1 -\ell }{\chi} \right)\! \eta_A(\chi) -  \left| \eta_B(\chi)\right|^2 \! \eta_B(\chi) = -    \eta_B(\chi)  \, .  \label{dimensionless2} 
\end{eqnarray}
Here the dependence of the solution on the choice of angular quantum number $\ell$ is implied. The asymptotic form of these equations shows that moduli for convergent solutions $|\eta_{A(B)}|$ can approach either $0$ or $1$ for large $\chi$. To study the analytic structure as well as a practical starting point for application of numerical methods, we expand the solution in a power series in $\chi$
\begin{eqnarray}
\eta_{A}(\chi) = \sum_{j=0}^\infty a_j \, \chi^j   \, , \hspace{2pc} \eta_{B}(\chi) = \sum_{j=0}^\infty b_j\,  \chi^j \,  . \label{NLDEseries}
\end{eqnarray}
Substituting these forms into the NLDE gives relations for the expansion coefficients
\begin{eqnarray}
&&\hspace{0pc} (1+\ell) b_1 - a_0^3 = - a_0  \,  ,           \\
&&\hspace{0pc}(2 +\ell) b_2 -3 a_1 a_0^2 =  - a_1 \, ,  \\
&&\hspace{0pc}  (3 +\ell) b_3  - 3 a_2 a_0^2 - 3 a_0 a_1^2 =  -  a_2   \, ,  \\
&& \hspace{3pc} \vdots \nonumber \\
 &&\hspace{0pc} ( 2- \ell) a_1+ b_0^3  =  b_0 \, , \\
  && \hspace{0pc} ( 3- \ell) a_2 + 3 b_1 b_0^2 =  b_1\, ,  \\
  &&  ( 4- \ell) a_3 + 3 b_2 b_0^2 +3 b_0 b_1^2= b_2  \, ,  \\
  && \hspace{3pc} \vdots \; \hspace{3pc}.   \nonumber    
\end{eqnarray}
Choosing a particular value for $\ell$ determines the values of both $a_0$ and $b_0$ and we find that there is only one independent parameter. Thus, for a given value of $\ell$ a vortex is found by tuning $a_{\ell-1}$ towards a critical value $a_{\ell-1}^{\textrm{vortex}}$. As examples, we have found vortices for the three lowest $\ell$ values for which both spinor components have nonzero rotation
\begin{eqnarray}
a_1^{\textrm{vortex}}  =  0.571718...  \; \;, \hspace{3pc} \ell =2 \, ,  \\
a_2^{\textrm{vortex}}  =  0.145291...  \; \; , \hspace{3pc} \ell =3 \, ,\\
a_3^{\textrm{vortex}}  =  0.0240267...  \;\;  , \hspace{2.6pc} \ell =4   \, .
\end{eqnarray}
The associated radial profiles for these solutions are plotted in Fig.~\ref{highervortices}. In Fig.~\ref{NumVortexConv} we illustrate convergence for both spinor components for the $\ell =2$ vortex. Convergence to the vortex profile is seen by overlaying an excitation of the vortex (a) and the free-particle Bessel-like solution (b). Note that there are two types of radially excited solutions characterized by the property that $|\eta_{A,(B)}|$ oscillate around $1$ or around $0$. 

\begin{figure}[h]
\centering
\subfigure{
\label{fig:ex3-a}
  \includegraphics[width = .7 \textwidth]{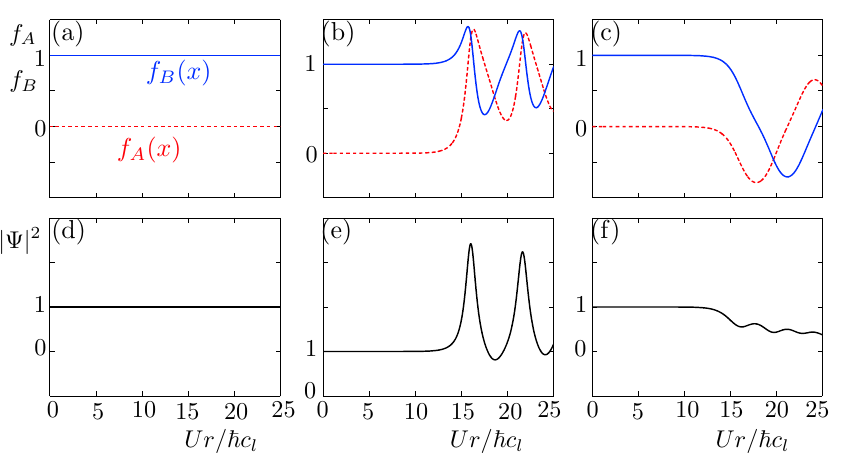}} \\
\caption[]{(color online) \emph{Radial excitations for $\ell=1$}. (a) Globally flat solution. (b) Strong nonlinearity. (c) Weak nonlinearity. }
\label{radial1}
\end{figure}

 We may elaborate further on the oscillating regimes on either side of the flat vortex profile by examining the solution for $\ell =1$. The $\ell = 1$ state corresponds to one unit of rotation in $\psi_B$ and no rotation in $\psi_A$, in which case a trivial constant solution exists, i.e., $\eta_A =1$ and $\eta_B =0$, for $0 \ge \chi > \infty$. We address this solution as it offers a clear look at higher radial excitations. To see this we solve Eqs.~(\ref{dimensionless1})-(\ref{dimensionless2}) by numerical shooting for $\ell =1$. The resulting spin component profiles are plotted in Fig.~\ref{radial1}(a)-(c) where we have also included plots of the associated total densities $|\Psi|^2 \equiv |\psi_A|^2 + |\psi_B|^2$ in Figs.~\ref{radial1}(d)-(f). Radial excited states appear as we tune the initial value for $\eta_A$, or equivalently $a_0$ in the Taylor expansion, away from unity. The spatially constant solution obtained for $a_0 =1$ can be thought of as a boundary between two oscillating regimes: one characterized by strong nonlinearity with oscillations around $\eta_A =1$, $\eta_B = 1$ in Fig.~\ref{radial1}(b); the other associated with weak nonlinearity with oscillations around $\eta_A = \eta_B =0$ shown in Fig.~\ref{radial1}(c). When tuning $a_0$ upward from unity, the total density in Fig.~\ref{radial1}(e) shows the onset of inward movement of bright vortex rings over a nonzero background. For this plot we have chosen $a_0 =  1 +  10^{-7}$. Here we see that the solution overshoots to an excited state of the vortex and enters a regime where the nonlinearity dominates the kinetic energy. In contrast, for the choice $a_0 =  1 - 10^{-7}$ Figs.~\ref{radial1}(c) and (f) show the onset of radial oscillations which result from a dominant kinetic energy in the equations of motion. In this case the total density exhibits decaying oscillations towards a zero background density. Here, undershooting the flat solution results in solutions which approach the Bessel function form as seen in Fig.~\ref{radial1}(c). This is what we expect to see for weak nonlinearity. This illustration highlights the property of the Dirac kinetic terms which act by confining the solution in the presence of strong nonlinearity.

\begin{figure}[t]
\centering
\subfigure{
\label{fig:ex3-a}
 \includegraphics[width=.7\textwidth]{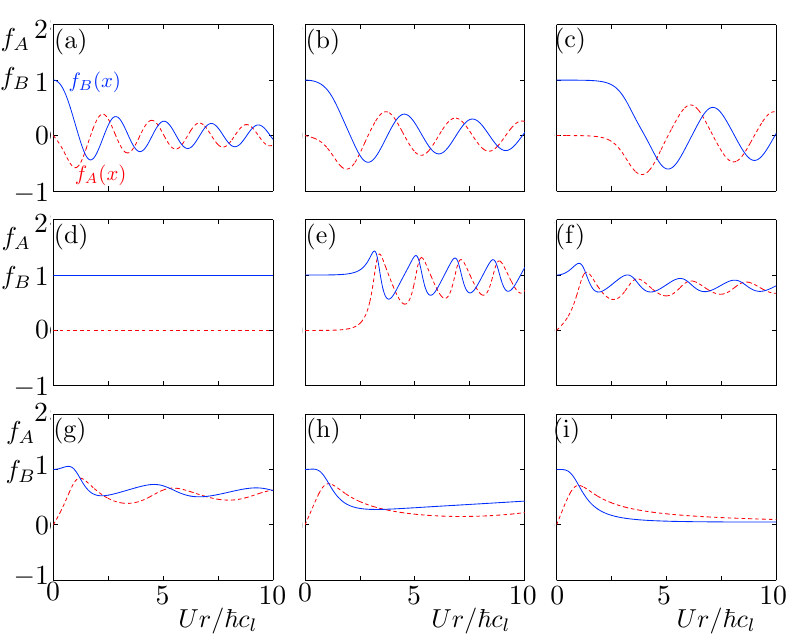}}  \\
\caption[]{(color online) \emph{Progression of radial profiles for vortices with phase winding $\ell =1$}. (a)-(c) Solutions where the renormalized chemical potential satisfies the condition $\tilde{\mu} > 1$. (d) Solution for $\tilde{\mu} = 1$. (e)-(i) Solutions for $\tilde{\mu} < 1$. The sequence of plots depicts the transition from the weakly-interacting Bessel-like solution in (a) for which $\mu > U$, to the strongly nonlinear case for which $\mu < U$ in (i). Note that $\tilde{\mu} =1$ is the boundary between solutions which oscillate around $0$ and solutions which oscillate around $1$. The results in these plots show that general ring-vortex solutions may be obtained by starting from excited states such as in (e), then reducing $\tilde{\mu}$ towards zero.}
\label{muzerosequence}
\end{figure}

So far in this section we have shown how tuning the initial condition above some critical value forces the solution into the strongly nonlinear regime, whereas below this value leads to solutions in the weakly nonlinear regime. In our analysis we held the chemical potential and interaction fixed, $\mu =U =1$. We now study the effect of tuning the chemical potential for $\mu > 1 \, \to \, \mu =0$, while maintaining the initial value in the shooting process and the interaction fixed. For arbitrary phase winding we will see a progression from Bessel solutions through a topological vortex, then finally ending in a single bright ring-vortex with vanishing tail. This solution resembles those found in attractive BECs and also in optics. Although we demonstrate such solutions by analytical methods in the next section, the numerical approach allows us to demonstrate the transition from the free-particle limit, $\mu \gg U$ (weak nonlinearity), to the vortex limit, $\mu \ll U$ (strong nonlinearity). To show this transition we choose a particular rescaling of the NLDE such that the limit $\mu \to 0$ makes sense, i.e., 
\begin{eqnarray}
\chi \equiv U^2 r/(\hbar c_l \mu ) \,  ,  \; \; \eta_A \equiv \sqrt{\mu/U} f_A \, , \; \;  \eta_B \equiv \sqrt{\mu/U} f_B \, .  \label{rescaling}
\end{eqnarray}
The resulting dimensionless NLDE is
\begin{eqnarray}
 -\left(\! \partial_{\chi}+\frac{\ell}{\chi} \right)\! \eta_B(\chi) +  \left|\eta_A(\chi)\right|^2\! \eta_A(\chi)     =   \tilde{\mu}^2 \,  \eta_A(\chi) \, ,  \label{dimensionless3}                       \\
  \left( \!\partial_{\chi}+ \frac{1 -\ell }{\chi} \right)\! \eta_A(\chi) +  \left| \eta_B(\chi)\right|^2 \! \eta_B(\chi)       =   \tilde{\mu}^2 \,  \eta_B(\chi)  \, .  \label{dimensionless4} 
\end{eqnarray}
 Starting with the case $\ell =1$, we fix $a_0 =1$ and $b_0 =0$ in the Taylor expansions, Eq.~(\ref{NLDEseries}), and tune $\tilde{\mu} \equiv \mu/U$ toward zero starting from $\tilde{\mu} > 1$. The progression of this solution as $\tilde{\mu} \to 0$ is depicted in the sequence of plots in Fig.~\ref{muzerosequence}. As $\tilde{\mu}$ is reduced towards $1$, Fig.~\ref{muzerosequence}(a)-(c), Bessel-like oscillations about $\eta_{A (B)} =  0$ are pushed out towards large $\chi$, completely flattening out the solution at $\tilde{\mu} =1$ in Fig.~\ref{muzerosequence}(d). As we continue decreasing $\tilde{\mu}$ towards $0$, oscillations about $\eta_{A (B)} =  1$ move inward from large $\chi$ and finally flatten out leaving only a coreless ring-vortex centered at $\chi=0$ shown in Fig.~\ref{muzerosequence}(i). The precise values of the renormalized chemical potential are $\tilde{\mu} = 1.22, \, 1.09, \, 1.0005, \, 1, \, 0.9995 , \, 0.775, \, 0.55 , \, 0.316, \, 0.1$, for the panels (a)-(i) in Fig.~\ref{muzerosequence}, respectively.

 \begin{figure}[t]
\centering
\subfigure{
\label{fig:ex3-a}
  \includegraphics[width=.7\textwidth]{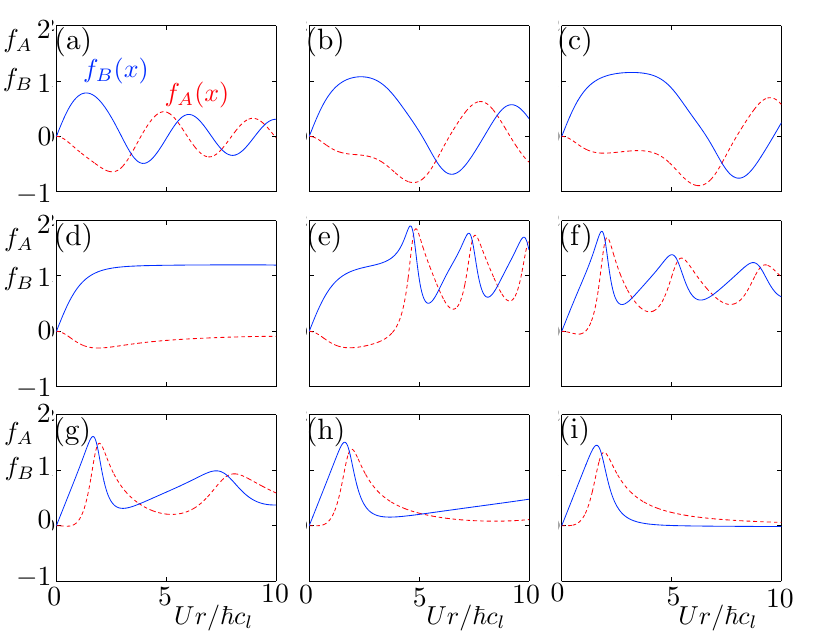}}  \\
\caption[]{(color online) \emph{Progression of radial profiles for vortices with phase winding $\ell =2$}. (a) $\tilde{\mu} =1.3$. (b) $\tilde{\mu} =1.21$. (c) $\tilde{\mu} =1.2047$. (d) $\tilde{\mu} =1.204267325$. (e) $\tilde{\mu} = 1.203$. (f)  $\tilde{\mu} = 0.85$. (g) $\tilde{\mu} = 0.55$. (h)  $\tilde{\mu} = 0.32$. (i) $\tilde{\mu} = 0.1 $. The topological vortex solution is shown in (d) and isolated ring-vortex in (i). }
\label{ringvortex2}
\end{figure}

 Analogous progressions are seen for arbitrary $\ell$ values culminating in a single ring-vortex centered at $\chi =0$ when $\tilde{\mu} =0$. The qualitative difference between the solution for $\ell = 1$ and those for $\ell >1$ is that in the later case both components must vanish at the origin. In terms of numerics this requires taking $b_0 = a_0 =0$, while specifying the first derivative of $\eta_A$ at the origin. The progression is plotted for the case $\ell =2$ in Fig.~\ref{ringvortex2}. Note that for each value of $\ell$ there are two distinct types of asymptotically flat vortices, as displayed in panels (d) and (i) of Fig.~\ref{ringvortex2}. The specific values of the renormalized chemical potential are $\tilde{\mu} = 1.3, \, 1.21, \, 1.2047, \, 1.204267325, \, 1.203, \, 0.85, \, 0.55, \, 0.32, \, 0.1$, for panels (a)-(i) of Fig.~\ref{ringvortex2}.

The NLDE allows for vortex solutions which satisfy an additional symmetry given by the constraint $|\eta_A|^2 + |\eta_B|^2 = 1$. Solutions which satisfy this constraint are the vortex/soliton or coreless vortex, Anderson-Toulouse and Mermin-Ho skyrmions. Such solutions may be obtained numerically by shooting backwards from large $\chi$ towards $\chi =0$. In particular, for the Mermin-Ho solution we integrate backwards starting with the asymptotic boundary condition $\eta_A = \textrm{cos}(\pi/4) + 10^{-k}$,  $\eta_B = \textrm{sin}(\pi/4) - 10^{-k}$. Here $k$ is a parameter tuned to give the desired values of functions at the origin, analogous to $a_0$ for the forward shooting. The Anderson-Toulouse vortex is obtained similarly but with the boundary condition $\eta_A = \textrm{cos}(\pi/2) + 10^{-k}$,  $\eta_B = \textrm{sin}(\pi/2) - 10^{-k}$. The half-quantum vortex is obtained by forming linear combinations of the numerical Mermin-Ho components, as previously discussed.

\section{Discrete Spectra in a Harmonic Trap}
\label{DiscreteSpectra}

\begin{figure}[h]
\centering
\subfigure{
\label{fig:ex3-a}
 \includegraphics[width=.7\textwidth]{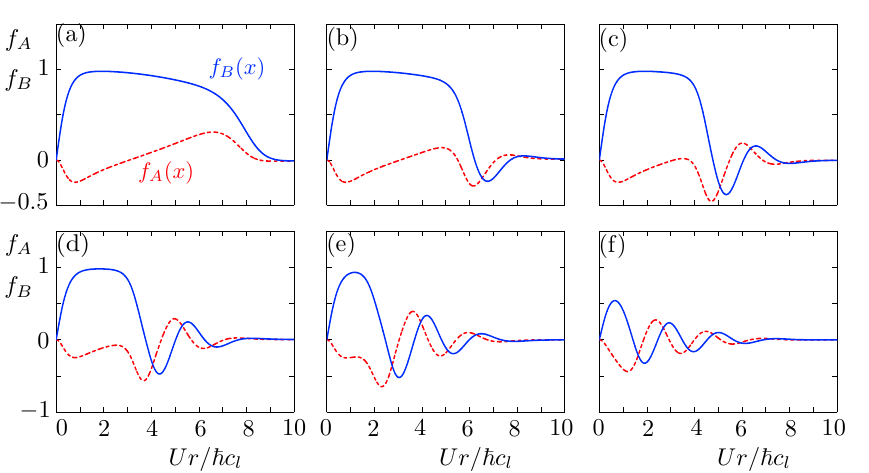}}  \\
\caption[]{(color online) \emph{Numerical shooting for quantized vortices in a quasi-2D harmonic potential}. (a)-(f) First six quantized states for $\ell =2$. The extreme vortex limit (a) is characterized by a dominant nonlinearity with the spinor functions flattening out over a greater portion of the domain with a weaker contribution from the derivate terms. In contrast, in the free-particle limit (f) spinor functions resemble Bessel oscillations.}
\label{cnfnd3}
\end{figure}

 We now extend our numerical studies from Sec.~\ref{NumericalVortices} to the treatment of vortices bound within a weak harmonic potential similar to the analysis in our previous work on solitons (a brief summery of these results can be found in~\cite{haddad2012}, presented here in full detail). Physically this accounts for an additional external harmonic trapping potential present in real experiments. We follow a similar procedure as before in preparing the NLDE for numerical analysis by first converting to a dimensionless form. Here we examine the same solutions as in Sec.~\ref{NumericalVortices} but with tight confinement assumed in one of three spatial dimensions and with additional weak harmonic confinement in the remaining two directions. The oscillator frequencies thus satisfy $\omega_z \gg \omega \equiv  \omega_x , \, \omega_y$. Equations.~(\ref{eqn:CondPsi7})-(\ref{eqn:CondPsi8}) are already defined for a quasi-2D system, as the z-dependence has been integrated out and parameters are normalized accordingly. Thus, we require the harmonic potential to be dependent only on the planar directions $x$ and $y$. We then introduce the planar-symmetric harmonic potential $V(r) = (1/2) M\, \omega^2 ( x^2 + y^2)^2 = (1/2) M \, \omega^2 r^2$. Next, we choose a dimensional rescaling of the NLDE appropriate to the harmonic oscillator. We divide through by the harmonic oscillator energy $\hbar \omega$ and define the dimensionless variable and spinor components in terms of this energy scale
\begin{eqnarray}
\chi \equiv \hbar \omega r /( \hbar c_l)\; , \;\;\; \eta_{A(B)} \equiv \sqrt{U/\hbar \omega} \, f_{A(B)}\, .
\end{eqnarray}
This transforms the NLDE to
\begin{eqnarray}
  -\left(\! \partial_{\chi}+\frac{\ell}{\chi} \right)\! \eta_B(\chi) +  \left|\eta_A(\chi)\right|^2\! \eta_A(\chi) + \mathcal{Q} \,  \chi^2 \, \eta_A(\chi)   =    \tilde{\mu} \,  \eta_A(\chi)  \, , \label{dimensionless3}                       \\
  \left( \!\partial_{\chi}+ \frac{1 -\ell }{\chi} \right)\! \eta_A(\chi) + \left| \eta_B(\chi)\right|^2 \! \eta_B(\chi) + \mathcal{Q} \,  \chi^2 \,  \eta_B(\chi)     =   \tilde{\mu} \,  \eta_B(\chi)  \, ,  \label{dimensionless4} 
\end{eqnarray}
where the two dimensionless parameters in our equations are
\begin{eqnarray}
 \mathcal{Q} \equiv  \frac{M c_l^2}{2\, \hbar \omega} \; , \;\;\;\; \;\;\; \tilde{\mu} \equiv \frac{\mu}{\hbar \omega} \, .
\end{eqnarray}
Solving Eqs.~(\ref{dimensionless3})-(\ref{dimensionless4}) subject to the radial boundary conditions consistent with the trap potential leads to radially quantized solutions labelled by integer quantum numbers. Fig.~\ref{cnfnd3} shows the first six quantized solutions for the $\ell=2$ topological vortex in panels (a)-(f), respectively. 

\begin{figure}[h]
\centering
\subfigure{
\label{fig:ex3-a}
 \includegraphics[width= .8\textwidth]{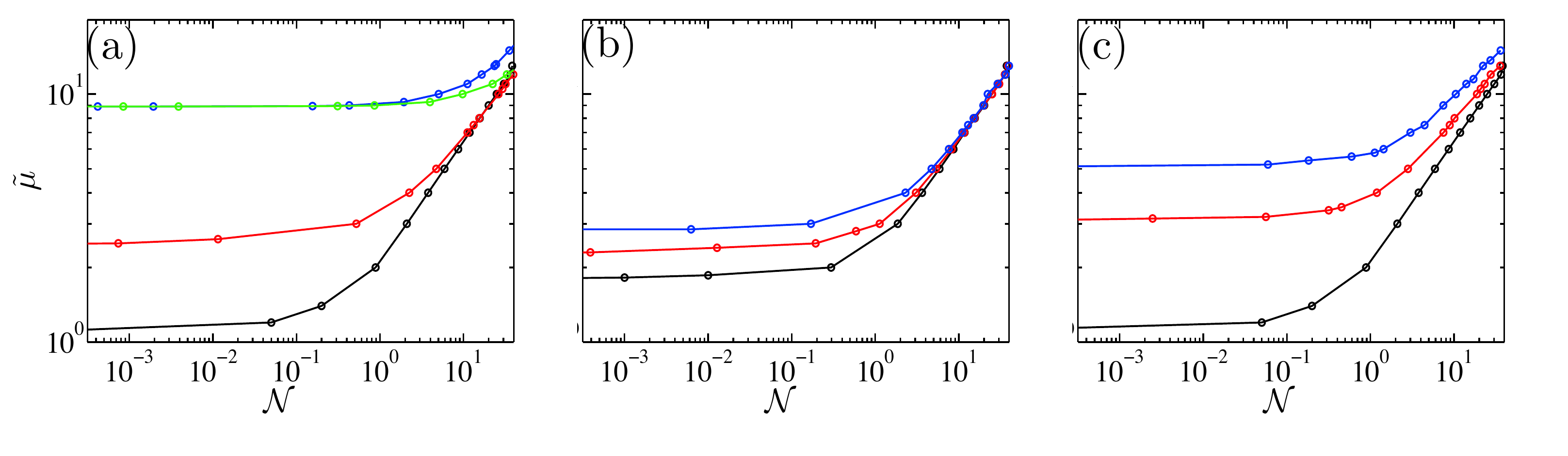}}  \\
\caption[]{(color online) \emph{Spectra for relativistic vortices confined in a harmonic potential}.(a) Vortex/soliton (black curve), Anderson-Toulouse skyrmion (red), Mermin-Ho skyrmion (blue), and half-quantum vortex (green). (b) General topological vortices for $\ell=2, 3, 4$ (black, red, blue). (c) Radial ground state and first two excited states of the vortex without skyrmion symmetry (black, red, blue). In each figure, the renormalized chemical potential is plotted as a function of the normalization.}
\label{completespectra}
\end{figure}

To connect our results to experiment we compute the discrete eigenvalue spectra for our radially quantized solutions which relate the chemical potential for a particular solution to the strength of the nonlinearity. The spectrum is computed by first defining the normalization condition as an integral over the dimensionless spinor components 
 \begin{eqnarray}
 \int \chi d\chi  ( |\eta_A(\chi)|^2 + |\eta_B(\chi)|^2) =  \mathcal{N} \, ,  \label{norm3}
\end{eqnarray}
where $\mathcal{N}$ on the right hand side encapsulates key lattice, condensate, and trap information through the relation
\begin{eqnarray}
\mathcal{N} = \frac{\sqrt{3}\,  \hbar \omega \, N  \, U}{ 3 \, t_h^2} \, .
\end{eqnarray}
The calculation proceeds by fixing the value of $\mathcal{Q}$ in Eqs.~(\ref{dimensionless3})-(\ref{dimensionless4}) and varying $\tilde{\mu}$, while computing the norm $\mathcal{N}$ for each value of $\tilde{\mu}$ by integrating Eq.~(\ref{norm3}). This procedure gives the paired values $(\mathcal{N}, \tilde{\mu})$. Fixing the total particle number $N$ gives a relation between the chemical potential $\mu$ and the interaction $U$. The values for the free parameter in the Taylor expansion, $a_0$, normalization $\mathcal{N}$, and corresponding chemical potential  $\tilde{\mu}= \mu/\hbar \omega$, are tabulated in Table~\ref{L2spectra} for the lowest radial $\ell =2$ mode shown in Fig.~\ref{cnfnd3}(a). Plots of this data along with the spectra for other solutions are shown in Fig.~\ref{completespectra}. We have taken $\mathcal{Q} = 0.001$ for all of our calculations. The spectrum for a fixed quantized mode tracks the flow of the chemical potential as one tunes the system between the free-particle and strongly nonlinear limits. In the former limit radial profiles for the lowest excitations resemble decaying Bessel functions as in harmonically trapped massless Dirac particles, whereas in the latter limit the nonlinearity dominates and the lowest trapped solutions flatten, consistent with the Thomas-Fermi regime. For higher quantized modes radial derivatives force the solution to maintain the Bessel-like form even as one tunes the parameters into the strongly nonlinear regime. Convergence of our solutions is provided in Appendix A.

\begin{table}[h]
\resizebox{12cm}{!}{
 \begin{tabular}{| c  | c | c  | }
\hline      
  \vspace*{-1mm}  &   \vspace*{-1mm}   &   \vspace*{-1mm}   \\
Free parameter $a_0$ &      Normalization $\mathcal{N}$ & Chemical potential $\tilde{\mu}$    \\
\hline  \vspace*{0mm}  
 $0.00000003$   & $1.9679 \times 10^{-5}$   &  $1.8$         \\
    \vspace*{-1.2mm} &  \vspace*{-1.2mm} &  \vspace*{-1.2mm} \\
 $0.0000 \, 0365$ &  $0.29560$   &  $2$         \\
        \vspace*{-1.2mm} &  \vspace*{-1.2mm} &  \vspace*{-1.2mm} \\
   $0.0000 \, 1018$  &  $1.85630$   &     $3$                \\
          \vspace*{-1.2mm} &  \vspace*{-1.2mm} &  \vspace*{-1.2mm} \\
    $0.0000 \, 1532 \, 51$  & $3.63288$    &     $4$       \\
          \vspace*{-1.2mm} &  \vspace*{-1.2mm} &  \vspace*{-1.2mm} \\
    $0.0000 \, 2089 \, 47$   &  $5.87547$   &    $5$     \\
         \vspace*{-1.2mm} &  \vspace*{-1.2mm} &  \vspace*{-1.2mm} \\
   $0.0000 \, 2708 \, 8395$  & $8.62326$    &     $6$       \\
          \vspace*{-1.2mm} &  \vspace*{-1.2mm} &  \vspace*{-1.2mm} \\
  $0.0000 \, 3387 \, 0293$    &  $11.72273$   &     $7$           \\
           \vspace*{-1.2mm} &  \vspace*{-1.2mm} &  \vspace*{-1.2mm} \\
   $0.0000 \, 4118 \, 5861 \, 75$   & $15.54209$   &   $8$           \\
              \vspace*{-1.2mm} &  \vspace*{-1.2mm} &  \vspace*{-1.2mm} \\
    $0.0000 \, 4899 \, 1422 \, 392$  &  $20.12707$   &    $9$     \\
                \vspace*{-1.2mm} &  \vspace*{-1.2mm} &  \vspace*{-1.2mm} \\
 $0.0000 \, 5725 \, 2724 \, 4133$  &  $24.87695$   &    $10$     \\
            \vspace*{-1.2mm} &  \vspace*{-1.2mm} &  \vspace*{-1.2mm} \\
   $0.0000 \, 6594 \, 2040 \, 6807 \, 59$    & $30.25986$   &       $11$     \\
                 \vspace*{-1.2mm} &  \vspace*{-1.2mm} &  \vspace*{-1.2mm} \\
    $0.0000 \, 75036 \, 2691 \, 0578 \, 11$     & $35.51507$   &   $12$       \\ 
              \vspace*{-1.2mm} &  \vspace*{-1.2mm} &  \vspace*{-1.2mm} \\
   $0.0000 \, 8451 \, 5722 \, 6783 \, 7827 \, 3795$        &  $37.30714$   &      $13$          \\ 
        \hline
\end{tabular}}
{\caption{\emph{Numbers for computing spectra of $\ell =2$ topological vortex \emph{ground state} solution.} For a fixed value of the chemical potential $\tilde{\mu}$, the free parameter $a_0$ is tuned until the desired radially quantized state is reached, for which one then computes the normalization $\mathcal{N}$ by Eq.~(\ref{norm3}). As $\tilde{\mu}$ increases into the Thomas-Fermi regime the dependence of the solution on $a_0$ becomes more sensitive, requiring a greater degree of tuned accuracy as shown in the column on the left.     }   \label{L2spectra}}
\end{table}

\section{Conclusion} 
\label{Conclusion}

In this article we have solved the NLDE by a variety of different methods for both the idealized case of zero trapping potential and in the harmonic case. The latter context provides the opportunity to study spatial quantization of vortex solutions in the radial dimension. A preliminary asymptotic solution using Bessel function expansions provides insight into the structure of the NLDE itself. Algebraic solutions were then obtained by considering the case of zero chemical potential. For these solutions the derivative and nonlinear terms are perfectly balanced leading to bright vortex rings over a zero-density background, a direct consequence of the Dirac operator. More generally, for nonzero chemical potential, we found that analytical solutions are only possible when one unit of winding is considered. In this case we used a numerical approach to obtain vortices with arbitrarily large winding number. A combination of numerical and analytical techniques yields skyrmion and half-quantum vortices, i.e., textures.

Having obtained our solutions, we computed their discrete spectra in the presence of a weak harmonic potential. This gives us the low-temperature $\mu$ versus $U$ landscape for relativistic vortices, where $\mu$ and $U$ are the chemical potential and lattice renormalized particle interaction, respectively. For example, for finite $\mu$ we found that a series of phase transitions occur as $U$ is tuned from zero upward: we encounter a Mermin-Ho skyrmion transitioning into a half-quantum vortex, followed by the Anderson-Toulouse skyrmion, then finally into a vortex/soliton, i.e., a bright soliton at the core of a singly-wound vortex.

Some of our solutions are similar to those obtained in spinor BECs, skyrmions in particular. However, presence of the Dirac operator contrasts heavily with the Laplacian case. This difference is most obvious in our bright ring-vortex solutions. Components of these solutions resemble the bright vortex which occurs in attractive BECs, but in our case the confining (focusing) regime of the Dirac operator is responsible for the effect in spite of repulsive atomic interactions. Furthermore, we should emphasize the significant distinction between our results and similar confinement effects in spinor BECs and in traditional models such as Thirring and Gross-Neveu. In each of these other theories the presence of confinement relies on attractive interactions in addition to a mass term, whereas we consider strictly the repulsive case with zero mass.

The interdisciplinary nature of our work suggests several future research directions. For instance, the form of our solutions implies possible mappings to finite energy solutions to classical gauge field equations. In particular, there is a deep connection between the NLDE and Chern-Simons terms relevant to quantum Hall fluids and more generally to relativistic field theories~\cite{Jackiw1990,Dunne1991,Horvathy2008}. To see this connection consider that Eq.~(\ref{compactNLDE}) resembles the low-energy effective theory for massless Dirac fermions interacting through a gauge field, where the latter has been absorbed into the local fermion contact terms. The Dirac term by itself is symmetric under global phase and chiral rotations which are made local through the addition of a gauge field. Quantization results in the well known axial anomaly which one finds to have exactly the Chern-Simons structure~\cite{Adler1969,Jackiw1969,Nakahara2003}. Thus, quantized theories of interacting fermions generally result in couplings to Chern-Simons terms. In our case we have focused only on the fermion part of the argument which naturally retains imprints of the omitted Chern-Simons terms. Another argument for the NLDE/Chern-Simons connection hinges on the well established duality between the Thirring model in (2+1)-dimensions and Maxwell-Chern-Simons theory. The mapping is arrived at through bosonization, i.e., reformulation in terms of paired particle and antiparticle field operators effective at strong coupling or low energy. These points merit further inquiry into potentially fruitful questions. Finally, in light of the successful analogs to date connecting condensed matter and nonlinear optics we expect that our results should be reproducible within an optics setting. Another obvious direction is to go deeper into the mathematics of nonlinear partial differential equations. The vast amount of work in this area provides an array of solution and classification methods which may be used to fully understand critical bounds for well-posedness as well as general solutions to the full NLDE with time dependence.

\ack{This material is based in part upon work supported by the National Science Foundation under grant numbers PHY-1067973, PHY-1011156, and the Air Force Office of Scientific Research grant number FA9550-08-1-0069. L.D.C. thanks the Alexander von Humboldt foundation and the Heidelberg Center for Quantum Dynamics for additional support.  }

\appendix

\section{Convergence of numerical vortex solutions}
\label{appendix}

\begin{figure}[h]
\centering
\hspace{-.075in}\subfigure{
\label{fig:ex3-b}
\includegraphics[width=.7\textwidth]{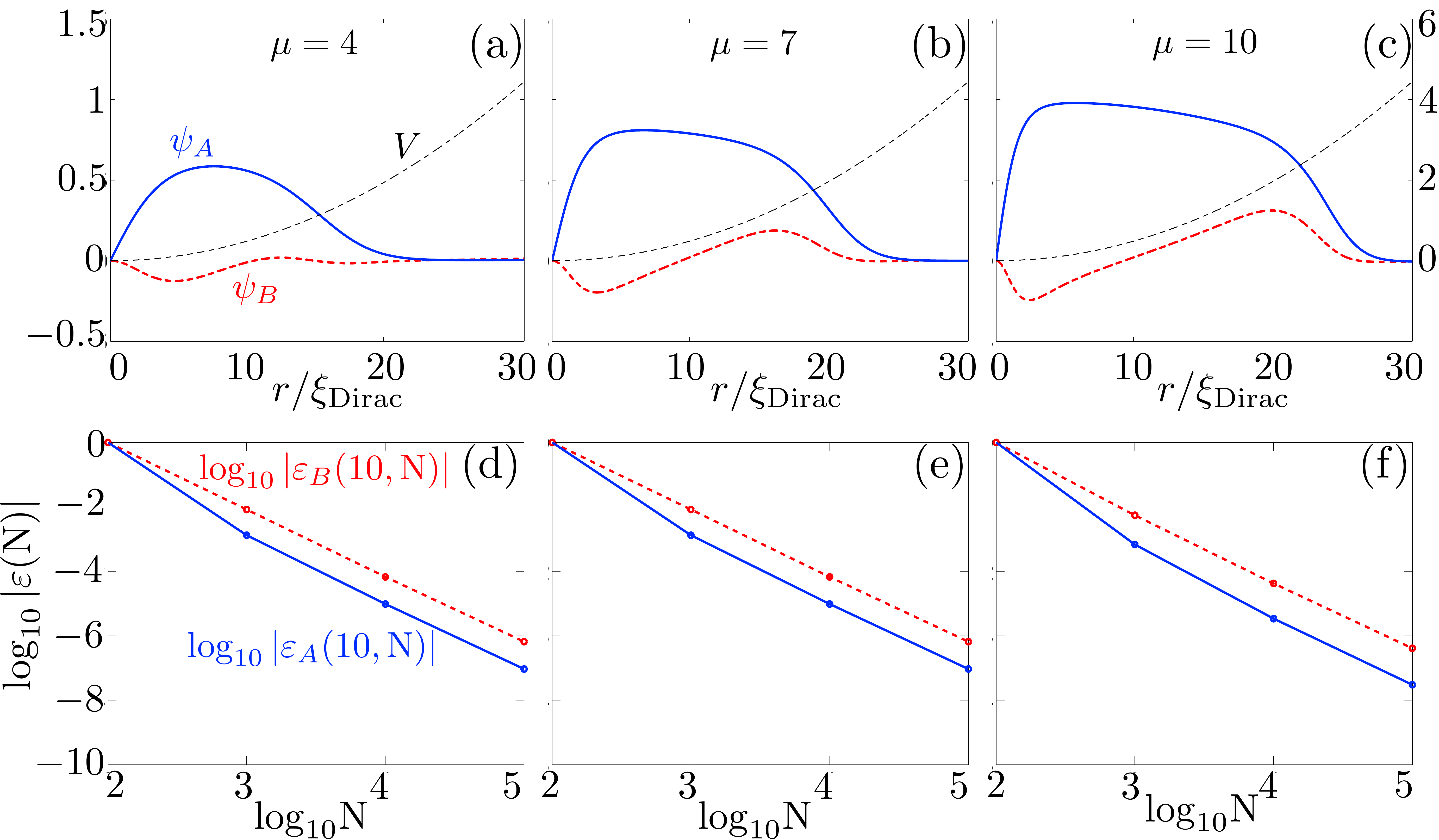}}\\
\caption[]{(color online) \emph{Convergence of $\ell =2$ topological vortex radial profiles}. (a)-(c) The explicit radial profiles for $U= 1$, $\mu =4, \, 7, \, 10$ and grid size $\mathrm{N} = 10^6$. The black dashed curve is the harmonic potential. The scale for the potential is shown on the right hand vertical axis of panel (c) in units of $10^{-4}    \mathrm{nK}$. (d)-(f) Log-log error profiles computed using Eq.~(\ref{error}). Note that the curves are a guide to the eye with data points representing actual data.  }
\label{VortWithConv}
\end{figure}

We demonstrate convergence of solutions in the harmonic trap for three of the $\ell =2$ topological vortices associated with the black curve in Fig.~\ref{completespectra}(b). Radial profiles for the chosen solutions are shown in Figs.~\ref{VortWithConv}(a)-(c). The corresponding values of the chemical potential are $\mu = 4$, $\mu = 7$, and $\mu = 10$. These values interpolate between the free-particle and strongly nonlinear limits (small to large $\mu$ values). These solutions were obtained by finite differencing using a shooting method to tune the precision of the initial value of $\psi_A$ such that $\psi_A \ll 1$ to pick out the ground state. For convergence at a single radial point, we compute the value of the solution at the dimensionless radius $\chi_i \equiv r_i/\xi_{\mathrm{Dirac}} = 10$ for several values of the grid size $\mathrm{N} = 10^2, \, 10^3, \, 10^4, \, 10^5, \,10^6$. We use the error formula which depends on the dimensionless radius and number of grid points 
\begin{eqnarray}
\varepsilon_{A(B)}(\chi_i, \mathrm{N})  \equiv \left[ \frac{  \psi(\chi_i)_{A(B)}^{\mathrm{N}+1} - \psi(\chi_i)_{A(B)}^\mathrm{N})}{ \psi(\chi_i)_{A(B)}^{\mathrm{N}+1} + \psi(\chi_i)_{A(B)}^\mathrm{N}) }\right] \, , \label{error}
\end{eqnarray}

\noindent where in the symbol $\psi(\chi_i)_{A(B)}^{\mathrm{N}}$ the subscript $A(B)$ denotes the sublattice excitation, $\chi_i$ denotes the $i^{\mathrm{th}}$ element in the discretized dimensionless radial coordinate, and the superscript $\mathrm{N}$ denotes the number of grid points used in the calculation. In Figs.~\ref{VortWithConv}(d)-(f), we have plotted $\mathrm{log}_{10}\left| \varepsilon_{A(B)}(\mathrm{10,N}) \right|$ versus $\mathrm{log}_{10}\mathrm{N}$, for the solutions shown in Figs.~\ref{VortWithConv}(a)-(c). 

\section*{References}

\bibliographystyle{unsrt}

\bibliography{NLDE_Vortex_Solutions_Refs}

\end{document}